\documentclass[aps,pre,superscriptaddress,twocolumn,floatfix]{revtex4}
\usepackage{graphicx}
\usepackage{amsmath}
\def\beq{\begin{equation}}
\def\eeq{\end{equation}}
\def\bea{\begin{eqnarray}}
\def\eea{\end{eqnarray}}

\def\beq{\begin{equation}}
\def\eeq{\end{equation}}
\def\bea{\begin{eqnarray}}
\def\eea{\end{eqnarray}}

\def\longrightharpoonup{\relbar\joinrel\rightharpoonup}
\def\longleftharpoondown{\leftharpoondown\joinrel\relbar}

\def\longrightleftharpoons{
  \mathop{
    \vcenter{
      \hbox{
      \ooalign{
        \raise1pt\hbox{$\longrightharpoonup\joinrel$}\crcr
	  \lower1pt\hbox{$\longleftharpoondown\joinrel$}
	  }
      }
    }
  }
}

\newcommand{\rates}[2]{\displaystyle
  \mathrel{\longrightleftharpoons^{#1\mathstrut}_{#2}}}

\begin{document}

\title{ Hidden long evolutionary memory in a model biochemical network
}

\author{Md. Zulfikar Ali}
\affiliation{Department of Physics, Clark University, Worcester, MA 01610}
\author{Ned S. Wingreen}
\email{wingreen@princeton.edu}
\affiliation{Department of Molecular Biology, Princeton University, Princeton, NJ 08540}
\author{Ranjan Mukhopadhyay}
\email{ranjan@clarku.edu}
\affiliation{Department of Physics, Clark University, Worcester, MA 01610}



\begin{abstract}
We introduce a minimal model for the evolution of functional protein-interaction networks using a sequence-based mutational algorithm, and apply the model to study neutral drift in networks that yield oscillatory dynamics. Starting with a functional core module, random evolutionary drift increases network complexity even in the absence of specific selective pressures. Surprisingly, we uncover a hidden order in sequence space that gives rise to long-term evolutionary memory, implying strong constraints on network evolution due to the topology of accessible sequence space.

\end{abstract}


\maketitle

Within even the simplest living cells there is a highly complex web of interacting molecules, with biological function typically emerging from the actions of a large number of different 
factors~\cite{Alberts, Alon}. What is the relationship between the architecture of such interaction networks and the underlying processes of evolution?  Much of the theory related to evolution focuses on the evolution of individual phenotypic traits or on population dynamics (see, for example, ~\cite{evolution1}); however, in general, individual genes do not determine individual traits. Rather, many traits arise from the dynamics of interacting components. With this in mind, we formulated and analyzed a minimal physically-based protein-protein interaction model that allows us to map from sequence space to interactions and, consequently, to network dynamics and fitness. Surprisingly, the model reveals a long-term memory of network origins hidden in the space of sequences.
 
 Recently, bottom-up approaches to molecular evolution, typically in the context of the folding properties/thermodynamics of individual proteins or RNAs~\cite{eigen,eigen2,Bloom,shakhnovich1,Zeldovich} have led to new insights into evolutionary outcomes, for example regarding a power-law distribution of protein family sizes. 
Here we generalize such bottom-up studies to functional networks. We focus on oscillatory networks of interacting enzymes, both due to the relevance of biological oscillators (e.g. cell cycle, circadian rhythms)~\cite{biooscillators, Circadian1, Circadian2} and due to the simplicity of defining function and fitness.  As such a network evolves, are the original nodes still both necessary and sufficient or does the network redistribute function over new nodes?
If new nodes do become essential, is there still
 memory of the original network?

   In order to address these questions, we develop a model of protein-protein interaction networks consisting of two classes of enzymes, activators (e.g. kinases) and deactivators (e.g. phosphatases). Each of these can be in either an active state or an inactive state and only function when in the active state. To model cooperativity, we assume that activation or deactivation of a target (either an activator or a deactivator) requires $h$ independent binding/modification events, with partially modified intermediates being short lived. The resulting chemical kinetic processes are
\bea
h {\rm A}^*_{i}+{\rm T}_{l} \xrightarrow{k_{il}} h{\rm A}^*_{i}+{\rm T}^*_{l} , \ \ 
h {\rm D}^*_{j}+{\rm T}^*_{l}  \xrightarrow{\tilde k_{jl}} h{\rm D}^*_{j}+{\rm T} _{l} ,
\label{Eqn1}
\eea
where A/A$^*$, D/D$^*$, and T/T$^* $ denote activator, deactivator, and target in inactive/active states respectively. The corresponding chemical kinetic equation can be approximated as (see Supplementary Material (SM)~\cite{suppm}, section I for details) 
\beq
\frac{d[{\rm T}^*_{l}]}{dt} = \sum^{m}_{i=1}k_{il}[{\rm A}^*_{i}]^h[{\rm T}_{l}] - \sum^{n}_{j=1} {\tilde k_{jl}}[{\rm D}^*_{j}]^h[{\rm T}^*_{l}]+\alpha[{\rm T}_{l}]- \alpha^{'}[{\rm T}^*_{l}] , 
\label{Eqn2}
\eeq
where $m$ and $n$ are the number of distinct types of activators and deactivators respectively. In Eq.~\ref{Eqn2}, $\alpha$ and $\alpha^{'}$ are background activation and deactivation rates. We further assume that the total concentration of each species is constant, such that ${\rm T}_{l} = c_0 - {\rm T}^*_{l}$.

Protein-protein interaction strengths are generally determined by amino-acid-residue interactions at specific molecular interfaces. Moreover, it has been estimated that $> 90\%$ of protein interaction interfaces are planar with the dominant contribution coming from hydrophobic interactions~\cite{shakhnovich, ProteinInteractions}. For simplicity, we therefore assume each protein possesses a pair of interaction interfaces, an in-face and an out-face, and we associate a binary sequence, ${\vec \sigma}_{\rm in/out}$, of hydrophobic residues (1s) and hydrophilic residues (0s) to each interface. The interaction strength between an enzyme (denoted by index $i$) and its target (denoted by index $l$) is determined by the interaction energy $E_{il}= \epsilon {\vec 
\sigma}_{\rm out}^{(i)} \cdot {\vec \sigma}_{\rm in}^{(l)} $ between the out-face of the enzyme and in-face of its target. (All energies are expressed in units of the thermal energy $k_{\rm B} T$.) The effective reaction rate is then given by 
\beq
k_{il} = k_{0} (1+ \exp [-(E_{il}-E_0)])^{-h} , 
\eeq
where $E_{0}$ plays the role of a threshold energy, e.g. accounting for the loss of entropy due to binding. The background activation and deactivation rates are set equal and define the unit of time via $\alpha =\alpha' = 1$. In our simulations we set $k_0 = 10^{4}$, $\epsilon = 0.2$, cooperativity $h=2$, $E_{0}= 5$, $c_0 = 1$, and we take the length of each sequence representing an interface to be $N=25$. These interaction parameters were chosen to provide a large range for the rate constants $k_{il}$ as a function of sequence and to keep the background rates small compared to the highest enzymatic rates; cooperativity was introduced to allow oscillations in relatively simple biomolecular networks. 

For our evolutionary scheme, we assume a population sufficiently small that each new mutation is either fixed or entirely lost~\cite{Moran, Nowak}. We consider only point mutations -- namely replacing a randomly chosen hydrophobic residue (1) in the in- or out-face of one enzyme by a hydrophilic residue (0), or vice versa. In this study, mutations are accepted if and only if they satisfy the selection criterion that the network remains oscillatory and moreover that the network exhibits oscillatory dynamics independent of the choice of initial concentrations of the active fractions (global oscillators). For this purpose we identified the fixed points of the chemical dynamics and carried out linear stability analysis (SM~\cite{suppm}, section II).

In order to address the question of network drift -- how function redistributes over the nodes in an evolving network -- we start with a 2-component oscillator (one activator and one deactivator) and add a second activator with all 0s for the sequences representing  in- and out-interfaces (so that initially Activator 2 has minimal interaction with the other two components). We then let the
system evolve, accepting only mutations corresponding to global oscillators. 
To characterize network drift, we studied the time evolution of the essentiality of each activator for a random sample of starting sequences that corresponded to oscillators, as depicted in Fig.~1A, where we characterize a component as being ``essential" if the 
system stops oscillating when the component is removed~\cite{footnote1}. In Fig.~1B we exhibit the distribution of the number of accepted mutational steps before the second activator become essential for  two distinct starting sequences. While the two distributions peak at very different values for the number of mutational steps, the interaction strengths for the two initial states do not differ appreciably (Fig.~1B, inset), highlighting the importance of the underlying sequence in governing evolutionary dynamics. Returning to Fig.~1A, we find relatively rapid flips between states where both activators are essential to states where only one of the activators is essential. 

\begin{figure}
\includegraphics[width = 8.5 cm, angle = 0]{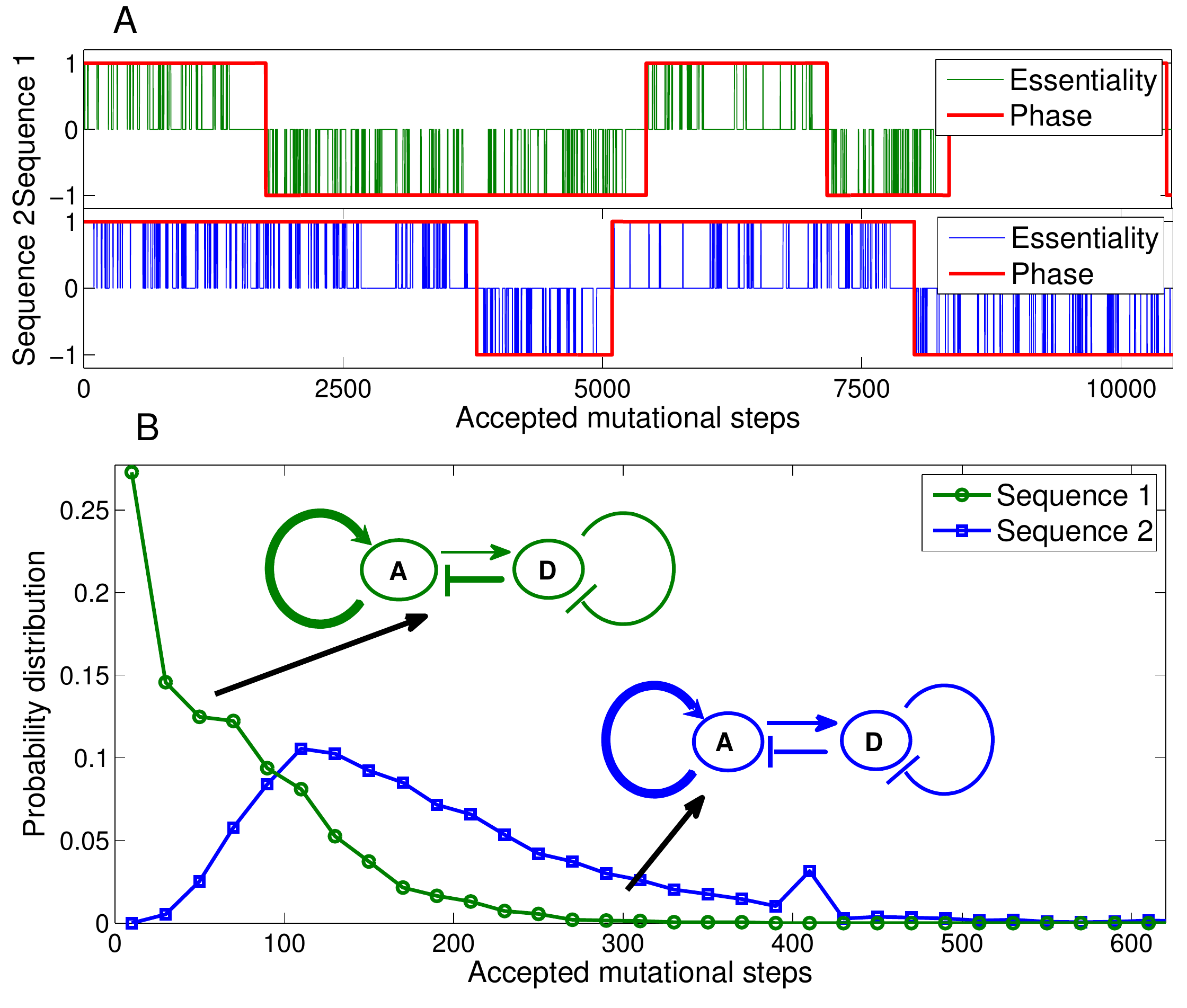}
\caption{Temporal evolution of essentiality of activators in 3-component systems. (A) Temporal evolution for two different initial sequences (the two sequences are specified in Supplementary Material (SM)~\cite{suppm}). On the $y$-axis, +1 indicates only Activator 1 is essential, -1 indicates only Activator 2 is essential, and 0 indicates both activators are essential. (B) Histograms of the number of accepted mutational steps before Activator 2 first becomes essential, for the two distinct initial sequences. Inset: interaction strengths of the two initial states. }
\end{figure}

Surprisingly, we also note the prevalence of much longer time periods where Activator 1 is always essential or where Activator 2 is always essential. This is true independent of initial conditions. These long evolutionary periods presumably reflect the division of sequence space into two regions or ``phases": Phase 1 where Activator 1 is always essential and Phase 2 where Activator 2 is always essential. The system starts in Phase 1 (Activator 2 is inessential), then when Activator 1 first become inessential we infer that the system has entered Phase 2, and so on. 

Can these two phases be distinguished in terms of measurable dynamical quantities or rate constants? Since the two phases presumably relate to an asymmetry in the roles of the two activators, we quantify this asymmetry via the relative peak-to-valley ratio (PVR) of the oscillations of their active fractions, where relative PVR is ((PVR A$_1$ - PVR A$_2$)/(PVR A$_1$ + PVR A$_2$)). From Fig.~2A (top panel) and Fig.~2B, we see that relative PVR correlates with the phase, and we display the distribution quantifying this correlation. A corollary is that the probability that an activator is essential also correlates with the relative PVR (Fig.~2C), so that if an activator has a relatively larger PVR it is also more likely to be essential. Moreover, we find that the phase-shift between peaks in the active fractions of the two activators also correlates with the phase (Fig.~2D), so that Activator 1 typically leads in Phase 1 and Activator 2 in Phase 2. Finally in order to determine how these observations relate to the underlying rate constants, we constructed the covariance matrix for the covariation of the 9 rate constants 
$k_{ij}$ and carried out a principal component analysis (SM~\cite{suppm}, section IV). We find that the projected component of the rates onto the eigenvector with the largest eigenvalue (PC1 = 94.93$\%$) strongly correlates with the phase (Fig.~2A, lowest panel, and Fig.~2E); we find no such correlation for projections onto any of the remaining eigenvectors. On examining the top eigenvector, we find that it primarily consists of a linear 
superposition of the difference in auto-activation rates of the two activators and the difference in their deactivation rates. This suggests that strong auto-activation coupled with strong deactivation produces an activator that peaks first during each oscillation cycle and also has a large PVR (SM~\cite{suppm}, section VIII). However, the co-occurrence of these features does not by itself explain the observed long intervals of two distinct phases.

\begin{figure}
\includegraphics[width = 8 cm, angle = 0]{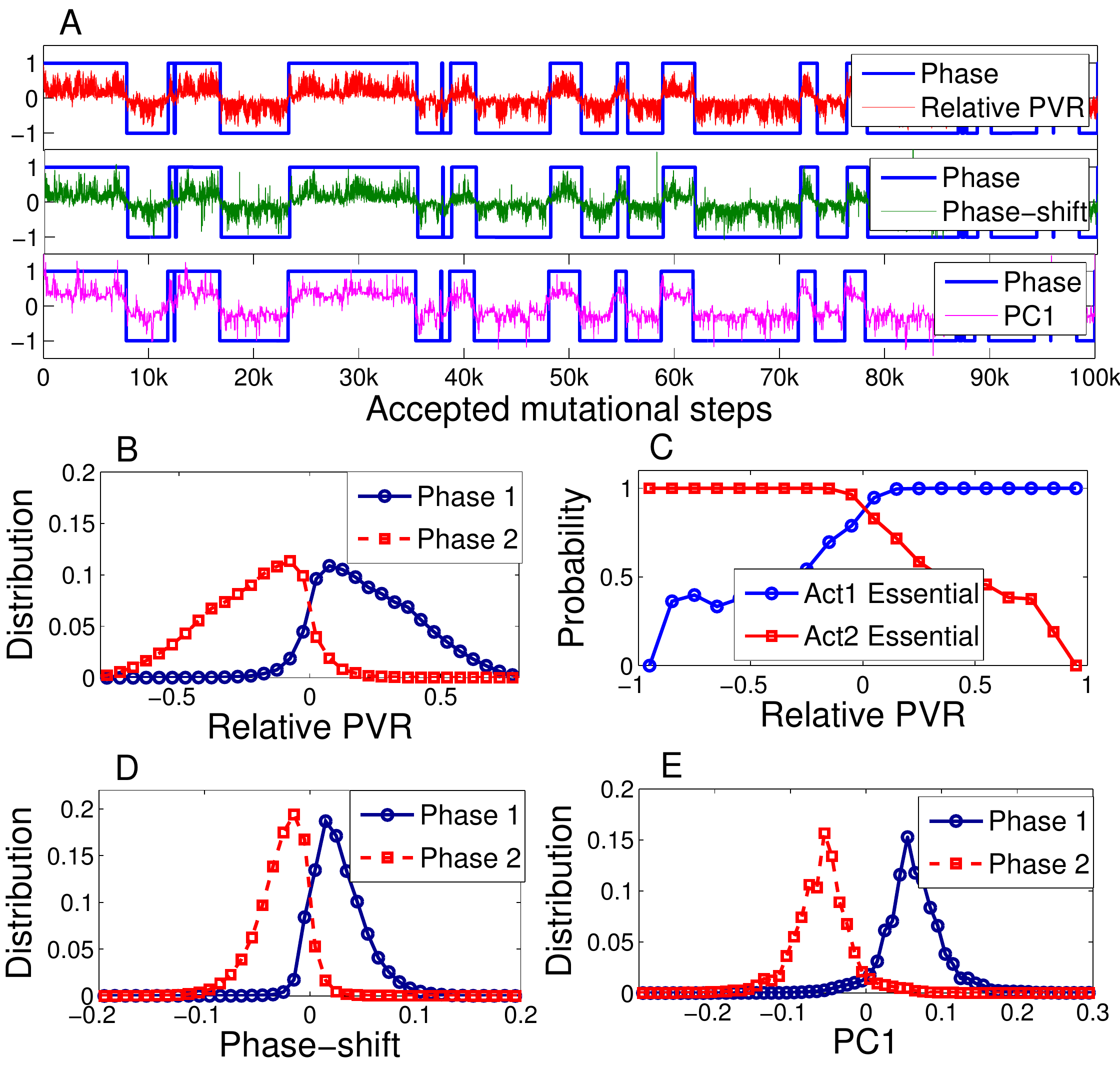}
\caption{Temporal evolution of phases in 3-component system. (A) Depiction of the temporal evolution where a value of +1 indicates Phase 1 and -1 indicates Phase 2. Along with the phase, the three panels show (i) normalized relative PVR of the two activators (red), (ii) phase-shift between their oscillatory peaks (green), and (iii) projected component of the chemical rates on the principal eigenvector from PCA analysis (magenta). (B) Distributions of relative PVR of the two activators in Phase 1 and in Phase 2. (C) Probability that each activator is essential as a function of its relative PVR. (D) Distribution of phase-shifts between active fraction peaks of the two activators in Phase 1 and Phase 2. (E) Distribution of projected rate constants on the principal eigenvector, obtained from PCA analysis, in Phase 1 and Phase 2.}
\end{figure}

What is the origin of the long-term memory? We first quantify the duration of long-term network memory by constructing a histogram of the number of mutational steps that the system spends in each phase before flipping. As shown in Fig.~3A, we find an approximately exponential distribution, $P(\tau) \propto e^{- \tau/\tau_{0}}$, where $\tau_{0} \simeq 3200 \pm 48 $ mutational steps. An exponential distribution implies a fixed, history-independent rate of flipping between the two phases, which in turn suggests that flipping corresponds to barrier crossing. Since our model treats all oscillatory states as equally fit, the only barriers are entropic, i.e. there must be relatively speaking very few boundary points connecting phases (SM~\cite{suppm}, section V). To check this hypothesis, we studied the neighborhood of states in Phase 1 and Phase 2. In Phase 1, for example, we distinguished between states where only Activator 1 is essential and states where both are essential. For states where only Activator 1 is essential we found no examples of sequences that were Hamming distance 1 away (that is, separated by a single point mutation) for which Activator 1 stops being essential. Of the states in Phase 1 where both activators are essential, for only $3 \%$ of states the Hamming distance 1 neighborhood contained one or more states where Activator 1 was inessential. The relative rarity of such states (which can be considered as boundary states) is consistent with our hypothesis that in sequence space the two phases touch at a relatively small number of boundary points. 

\begin{figure}
\includegraphics[width = 9 cm, angle = 0]{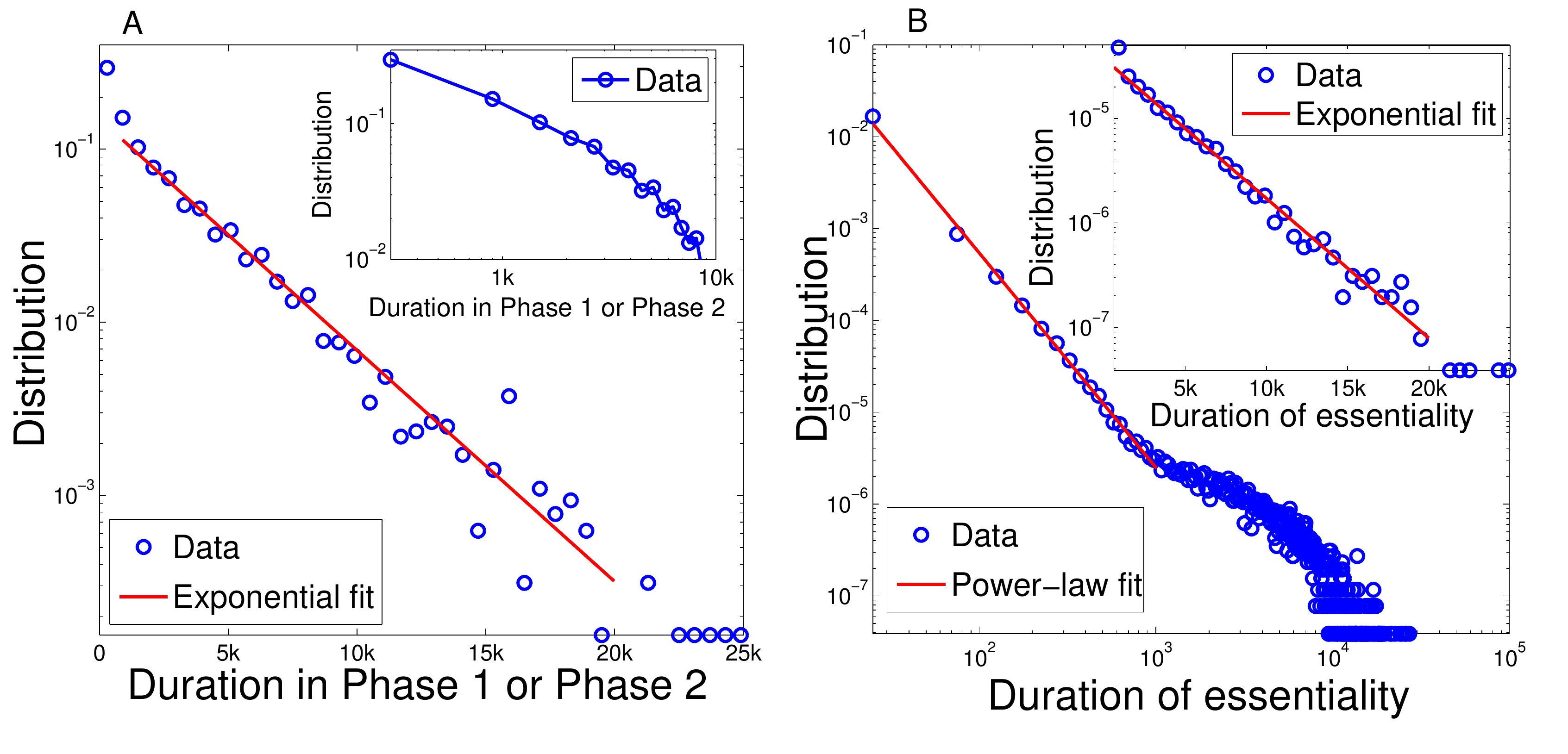}
\caption{Distribution of accepted mutational steps between flips. (A) Distribution of the number of accepted mutational steps between flips from one phase to the other, on a semi-log scale to highlight the exponential distribution (data is binned with bin size 600). Inset: same data on log-log scale. (B) Distribution of the number of accepted mutational steps where an activator is essential for the whole duration, on a log-log scale showing a power-law fit $f(x) \sim x^{-2.3\pm 0.05}$ for short times (bin size 50). Inset: same distribution over longer times on semi-log scale (bin size of 600).}
\end{figure} 

Interestingly, in contrast to flipping between phases, the distribution of the number of mutational steps that an activator remains essential exhibits a power-law distribution for short times, as depicted in Fig.~3B. For Activator 1, for example, this power-law part of the distribution is dominated by cases where the system is in Phase 2, with Activator 1 switching between being essential and inessential. Thus the power-law distribution is related to the presence of domains within Phase 2 where Activator 1 is also essential (and likewise for Activator 2 in Phase 1). For longer times, the periods of essentiality correspond to the duration of phases, and thus the distribution decays exponentially (Fig.~4B, inset). In contrast to exponential decay, a power-law distribution implies a history-dependent switching rate, with the escape rate from a domain proportional (on average) to the inverse of the time elapsed since the system entered the domain (SM~\cite{suppm}, section IX).

It is not {\it a priori} obvious how the above observations of two phases generalize to more complex networks. We therefore extended our study by starting with a 3-component oscillator and adding a fourth component (Activator 3) with all its sequences initially set to 0s. Once again we find that Activator 3 becomes essential relatively rapidly (typically in $\sim$100 mutational steps). If we continue to follow the evolution of essentiality for the activators, we find for each activator long periods ($\sim$1000+ mutational steps) where that activator remains essential, separated by similarly long periods where that activator is intermittently essential/inessential (Fig.~4A). This suggests that for each activator, the sequence space of oscillators divides into two regions: one region where that activator is essential at every point and a second region consisting of smaller domains where the activator is essential interspersed with domains where it is inessential. Note that time periods where one activator remains essential sometimes overlap with periods where one of the other activators remains essential, implying that the region where one activator is essential at every point has some overlap with the regions where other activators are essential at every point. This contrasts somewhat with the 3-component system where Phase 1, the region in which Activator 1 is essential at every point, is complementary to Phase 2. By contrast, as shown in Fig.~4B, the distribution of mutational steps over which any one of the activators is essential for the 4-component system is quite similar to that of the 3-component system, being power-law at short times with a similar exponent, and exponential for longer times, albeit with a shorter decay time $\tau_{0} \simeq 1750\pm 54$ mutational steps. As for 3-component systems, we also find strong correlation between normalized/relative PVR of oscillation, phase-shift, and essentiality for pairs of activators. We find that when the normalized PVR of an activator is higher, the probability that it is essential is also higher (Figs.~4C and 4D); these results generalize to much larger systems of activators and deactivators (SM~\cite{suppm}, section X). 

\begin{figure}
\includegraphics[width = 9 cm, angle = 0]{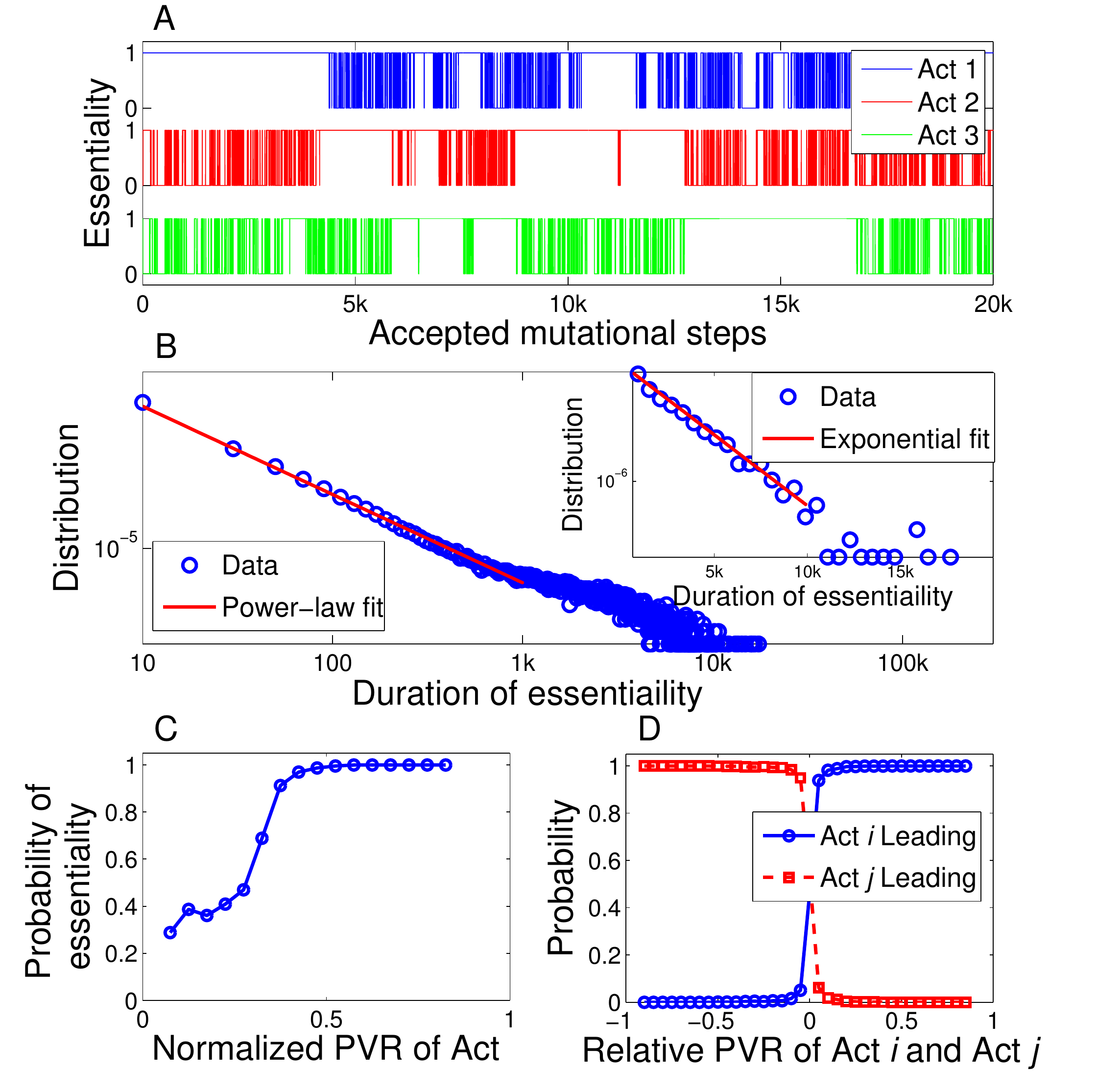}
\caption{Temporal evolution of essentiality of each activator in 4-component systems. (A) Depiction of temporal evolution where, on the $y$-axis, +1 indicates that the activator is essential and 0 indicates that it is not essential. (B) Distribution of the number of accepted mutational steps where the activator is essential for the whole period, on a log-log scale showing the power-law distribution $f(x) \sim x^{-2.15\pm0.02}$ for short times (bin size 20). Inset: same distribution on a semi-log scale. (C) Probability of Activator $i$ being essential as a function of its normalized PVR defined as PVR A$_i$/(PVR A$_1$ + PVR A$_2$ + PVR A$_3$). (D) For any pair of activators, the probability that Activator $i$ leads Activator $j$ as a function of their relative PVR. }
\end{figure}

In this paper, we focused on oscillatory networks and found that for a sequence-based scheme, evolution explores the space of possible oscillators in a manner strikingly different from in parameter-based evolution (see, for example, \cite{Siggia}). We studied how function can become distributed over new nodes due to random network drift. For a 3-node network, the typical timescale for the new node to become essential for oscillation is $\sim$100 point accepted mutations, which, given the total of 150 sites, corresponds to around $66 \%$ accepted mutations~\cite{PAM}. Surprisingly, our model also revealed a much longer term memory (around 2000 point accepted mutations per 150 amino acids for a 3-node system) with exponential decay, indicative of  a barrier crossing process in the space of sequences. 

We expect our model to be broadly useful for exploring principles of protein network evolution. While simple and easy to implement, the model is biologically grounded in sequence-based evolution, and also physically grounded insofar as all proteins interact via binding with all others. Within this approach, network topology emerges from evolutionary dynamics rather than being put in by hand. Moreover, there is no fine tuning and the degree of cooperativity utilized for the studies in this paper is modest and easily achievable in practice by biochemical networks~\cite{Ferrell}. The model provides a natural framework to study the interplay between selection pressure and sequence-based designability/accessibility. It can moreover be readily extended to larger networks, networks with other functions, and also to other mutation-selection regimes (for example, the concurrent mutations regime expected for larger populations~\cite{Desai}). 

We also believe our results for network drift will apply beyond the context of oscillators studied here. It has been suggested that protein networks evolve primarily by two biological mechanisms: (i) gene duplication, and (ii) random mutations in proteins leading to neo-functionalization, that is, the {\it de novo} creation of new relationships with other proteins~\cite{Dill}. Our studies illustrate the significance of neo-functionalization in the context of functional networks where protein-protein interactions are physically grounded, i.e. described via quantitative interaction strengths rather than Boolean variables. Our discovery of hidden order in sequence space leading to evolutionary long-term memory could also be quite general, highlighting the strong constraints to network evolution that emerge from the topology of accessible sequence space. It will be interesting to see if the presence of ``phases" generalizes to other network types. 
Future studies may profitably include the evolutionary dynamics of nodes, address other network functions (e.g. signal integration), and explore the role of graded selection in the {\it de novo} evolution of new functions.

  We acknowledge helpful discussions with Yigal Meir and Ammar Tareen. The research was supported in part by DARPA Biochronicity program, Grant D12AP00025, National Science Foundation Grant PHY-1305525, and National Institutes of Health Grant R01 GM082938.

\pagebreak
\widetext
\clearpage
\begin{center}
\textbf{\large Supplementary Materials: Hidden long evolutionary memory in model biochemical network}
\end{center}
\setcounter{equation}{0}
\setcounter{figure}{0}
\setcounter{table}{0}
\setcounter{page}{1}
\makeatletter

\newcommand\NT{\addtocounter{equation}{1}\tag{S\theequation}}
\newcommand{\R}[1]{\textrm{#1}}

\section{Chemical rate equations for system of interacting protein species}

We  consider a system consisting of $m$ species of activators (e.g. kinases), denoted by letter A, and $n$ species of deactivators (e.g. phosphatases), denoted by letter D, which can be in active or inactive states. Activators (in their active state) act only on inactive targets and deactivators (in their active state) act only on active targets. The chemical kinetic equations governing the system are given by

\begin{align}
\R{A}^*_{i}+\R{T}_{l} \rates{k^{f}_{il}}{k^{r}_{il}} \R{A}^*_i \R{T}_{l} \xrightarrow{r_{il}} \R{A}^*_{i}+\R{T}^*_{l}\NT \\
\R{D}^*_{j}+\R{T}^*_{l} \rates{{\tilde k}^{f}_{jl}}{{\tilde k}^{r}_{jl}} \R{D}^*_j \R{T}^*_{l} \xrightarrow{{\tilde r}_{jl}} \R{D}^*_{j}+\R{T}_{l}\NT 
\end{align}
where $\R{A}^*$, $\R{D}^*$ and $\R{T/T}^* $ denote activators, deactivators, and targets respectively ( $^{*}$ denotes active state). We assume that the sum of the active fraction and inactive fraction is constant and set it to 1. The rate of change of concentrations of the active fraction of the target, an activator, is given by
\begin{align}
\frac{d[\R{T}^*_{l}]}{dt} = \sum^{m}_{i\neq l}r_{il}[\R{A}^*_{i} \R{T}_{l}] + \sum^{n}_{j=1} \big({\tilde k}^r_{jl}[\R{D}^*_{j}\R{T}^*_{l}] - {\tilde k}^{f}_{jl}[\R{D}^*_{j}][\R{T}^*_{l}]\big ) + \bigg ( \big (2r_{ll}+k^r_{ll}\big )[\R{T}^*_{l} \R{T}_{l}] - k^f_{ll}[\R T^*_l][\R T_l] \bigg )  \nonumber \\
+ \sum^{m}_{i\neq l}\bigg ( \big (r_{li}+k^r_{li} \big )[\R{T}^*_{l} \R{A}_{i}] - k^f_{li}[\R T^*_l][\R A_i] \bigg ) +  \sum^{n}_{j=1}\bigg (\big (r_{lj}+k^r_{lj}\big )[\R{T}^*_{l} \R{D}_{j}] - k^f_{lj}[\R T^*_l][\R D_j] \bigg ),\NT
\end{align}
and for the active fraction of the target, a deactivator, is given by
\begin{align}
\frac{d[\R{T}^*_{l}]}{dt}  = \sum^{m}_{i = 1}r_{il}[\R{A}^*_{i} \R{T}_{l}] + \sum^{n}_{j\neq l} \big ({\tilde k}^r_{jl}[\R{D}^*_{j}\R{T}^*_{l}] - {\tilde k}^{f}_{jl}[\R{D}^*_{j}][\R{T}^*_{l}] \big ) + \bigg (\big ({\tilde r}_{ll}+2{\tilde k}^r_{ll} \big )[\R{T}^*_{l} \R{T}^*_{l}] - 2{\tilde k}^f_{ll}[\R T^*_l]^2 \bigg )  \nonumber \\
+ \sum^{m}_{i = 1}\bigg ( \big ({\tilde r}_{li}+{\tilde k}^r_{li} \big )[\R{T}^*_{l} \R{A}_{i}] - {\tilde k}^f_{li}[\R T^*_l][\R A_i] \bigg ) +  \sum^{n}_{j \neq l}\bigg ( \big ({\tilde r}_{lj}+{\tilde k}^r_{lj} \big )[\R{T}^*_{l} \R D^*_{j}] - {\tilde k}^f_{lj}[\R T^*_l][\R D^*_j] \bigg ) \NT.
\end{align}
The last three terms in Eqs. S3 and S4 are the terms when the target itself acts as an enzyme. For inactive fractions of the target, both activator and deactivator, the rate of change of concentrations is given by
\begin{align}
\frac{d[\R{T}_{l}]}{dt} = \sum^{m}_{i=1} (k^r_{il}[\R{A}^*_{i}\R{T}_{l}] - k^{f}_{il}[\R{A}^*_{i}][\R{T}_{l}]) + \sum^{n}_{j=1}{\tilde r}_{jl}[\R{D}^*_{j}\R{T}^*_{l}]  \NT,
\end{align}
and for each intermediate complex, the rate of change of concentration is given by
\begin{align}
\frac{d[\R{A}^*_{i}\R{T}_{l}]}{dt} &= k^{f}_{il}[\R{A}^*_{i}] [\R{T}_{l}]-(k^{r}_{il}+r_{il})[\R{A}^*_{i}\R{T}_{l}]  \NT \\
\frac{d[\R{D}^*_{j}\R{T}^*_{l}]}{dt} &= {\tilde k}^{f}_{jl}[\R{D}^*_{j}][\R{T}^*_{l}]-({\tilde k}^{r}_{jl}+ {\tilde r}_{jl})[\R{D}^*_{j} \R{T}^*_{l}]\NT .
\end{align}

Under the assumption that the intermediate complex concentrations are at steady state (quasi-static approximation), we obtain
\begin{align}
k^{f}_{il}[\R{A}^*_{i}][\R{T}_{l}]-(k^{r}_{il}+r_{il})[\R{A}^*_{i}\R{T}_{l}] &= 0\nonumber \\
{[\R{A}^*_{i}\R{T}_{l}]}  &=  \frac{k^{f}_{il}}{k^{r}_{il}+r_{il}}[\R{A}^*_{i}][\R{T}_{l}]  \NT \\
{\tilde k}^{f}_{jl}[\R{D}^*_{j}][\R{T}^*_{l}]-({\tilde k}^{r}_{jl}+r_{jl})[\R{D}^*_{j} \R{T}^*_{l}] & =  0\nonumber \\
{[\R{D}^*_{j}\R{T}^*_{l}]} & =  \frac{{\tilde k}^{f}_{jl}}{({\tilde k}^{r}_{jl}+ {\tilde r}_{jl})}[\R{D}^*_{j}][\R{T}^*_{l}] \NT.
\end{align}
Substituting Eqs. (S8) and (S9) in  Eqs. (S$3$) and (S$4$) yields same expression for the rate of change of concentration for the active fraction of target whether it is an activator or deactivator and is given by
\begin{align}
\frac{d[\R{T}^*_{l}]}{dt} &=  \sum^{m}_{i=1}k_{il}[\R{A}^*_{i}][\R{T}_{l}] - \sum^{n}_{j=1} {\tilde k}_{jl}[\R{D}^*_{j}][\R{T}^*_{l}] \NT .
\end{align}
Substituting Eqs. (S8) and  (S9) in Eq. (S$5$), we obtain
\begin{align}
\frac{d[\R{T}_{l}]}{dt} & = -\sum^{m}_{i=1}k_{il}[\R{A}^*_{i}][\R{T}_{l}] + \sum^{n}_{j=1} {\tilde k}_{jl}[\R{D}^*_{j}][\R{T}^*_{l}] \NT ,
\end{align}
where 
\begin{align}
k_{il} & =  \frac{k^{f}_{il}}{1 + k^{r}_{il}/r_{il}}, \nonumber \\
{\tilde k}_{jl} & = \frac{{\tilde k}^{f}_{jl}}{1 + {\tilde k}^{r}_{jl}/{\tilde r}_{jl}}  \NT .
\end{align}
As a check for self-consistency, we confirm that the rates of change of concentration of the active and inactive fractions are equal in magnitude but opposite in sign, as expected, since the sum of the concentrations of the active, inactive, and intermediate complexes is constant. This quasi-static approximation is justified in the limit where the concentrations of the intermediate complexes are small compared to the concentrations of the active and inactive fractions, that is, in the limit where the ratios $k_{il}^{f}/(k^{r}_{il} + r_{il})$ and ${\tilde k}_{il}^{f}/({\tilde k}^{r}_{il} + {\tilde r}_{il})$
are much smaller than 1, corresponding to relatively short-lived intermediate complexes. We also point out that this approximation neglects competition between targets due to sharing of enzymes \cite{Rowland}, which could become important for high affinity targets.
Under this simplifying assumption, we need only one rate equation for the active fraction of each species, thus reducing the number of rate equations to the number of enzyme species. Without loss of generality, we assume $k^{f}_{il}$ ($= {\tilde k}^{f}_{jl}$), and $r_{il}$ ($= {\tilde r}_{jl}$) are the same constants for all enzyme-target pairs, so that the only rate constants that depend on binding energies are $k^{r}_{il}$ and ${\tilde k}^{r}_{il}$. For these rates, we assume an Arrhenius-type form, e.g. 
$k^{r}_{il} = A e^{- E_{il}/k_\R{B} T}$ where $A$ is a constant and $E_{il}$ is the binding energy between enzyme represented by label $i$ and target represented by label $l$. If energy $E_{il}$ is measured in units of $k_\R{B}T$ where $T$ is room temperature, we obtain
 \beq
k_{il}  =  {k^{'}_{0}} \left( \frac {1} {1 + e^{-(E_{il} - E_{0})}} \right) \NT ,
\eeq 
where $k^{'}_{0}$ and $E_{0}$ are constants. 
Similarly, 
${\tilde k}_{jl}  =  k^{'}_{0}/(1 + e^{-({\tilde E}_{jl} - E_{0})})$. We have so far ignored background (enzyme-independent) activation and deactivation of target; we can incorporate this by adding  a term of the form $\alpha [\R{T}_{l}] - \alpha'[\R{T}^{*}_{l}]$ to the right-hand side of Eq. (S10). 

  We next incorporate cooperativity within our minimal model. For this purpose, we assume a two-stage (or more generally, $h$-stage) enzyme-mediated activation of target molecules with relatively short-lived intermediates. This could correspond, for example, to two phosphorylation sites for each target molecule, where both sites have to be phosphorylated for molecules to be active.  For simplicity of the discussion, we first consider a two-component system (1 activator, 1 deactivator species), and assume target molecules can be in three states: inactive (T), active (T$^{*}$), and partially phosphorylated (T$'$).  The  chemical kinetic equations governing the system are then of the form: 
\begin{align}
\R{A}^*+\R{T}  & \xrightarrow{k'}  \R{A}^* +\R{T}' \nonumber \\  
\R{T}'  & \xrightarrow{k'_{-}}  \R{T} \nonumber \\
\R{A}^*+\R{T}'  & \xrightarrow{k''}  \R{A}^* +\R{T}^{*}  \NT .
\end{align}
Within the assumption of short-lived complexes discussed above, the rates of change of T$'$ and T$^{*}$ are given by
\begin{align}
\frac{d[\R{T}']}{dt} &= k' [\R{A}^*][\R{T}] -  k'_{-}[\R{T}'] - k'' [\R{A}^*][\R{T}'] \nonumber \\
\frac{d[\R{T}^{*}]}{dt} &= k'' [\R{A}^*][\R{T}'] -  {\tilde k}[\R{D}^*][\R{T}^*] + \alpha [\R{T}] - \alpha'[\R{T}^{*}]  \NT .
\end{align}
Applying the quasi-static approximation for T$'$ (valid for low concentrations of [T$'$]), we obtain $ [\R{T}'] = k' [\R{A}^*][\R{T}]/(k'_{-} + 
k'' [\R{A}^*])$. If  $k'' [\R{A}^*] \ll k'_{-}$ (high spontaneous decay rate of partially phosphorylated state), we can further approximate $ [\R{T}'] \approx k' [\R{A}^*][\R{T}]/k'_{-} $. We thus obtain
\beq
\frac{d[\R{T}^{*}]}{dt} = k [\R{A}^*]^{2}[\R{T}] -  {\tilde k}[\R{D}^*][\R{T}^*] + \alpha [\R{T}] - \alpha'[\R{T}^{*}] \NT ,
\eeq
where $k = k' k''/k_{-}$. Both $k'$ and $k''$ can be expected to be of the form in Eq. (S15), while $k_{-}$ as a spontaneous decay rate can be treated as a constant. 
For simplicity, we assume that the enzyme binding energies for both steps of phosphorylation are the same, and obtain
\beq
k  =  {k_{0}} \left( \frac {1} {1 + e^{-(E - E_{0})}} \right)^{2} \NT ,
\eeq
where $k_{0}$ is now a new constant. Along similar lines, we can also introduce cooperativity in deactivation via a two-stage enzyme-mediated deactivation process. We then generalize this to multiple activator/deactivator species, with the simplifying assumption that activators/deactivators involved in both stages belong to the same species, giving us Eqs. (2) and (3) in the main text. 

\section{Linear stability analysis}
To check if the system corresponds to a global oscillator, we used linear stability analysis as a tool. We will describe it here for a two component system: this can be generalized for N-component systems. For 1 activator and 1 deactivator, the chemical rate equation is written as

\begin{align*}
\frac{d[\R{A}^*]}{dt} & = k_{AA}[\R{A}^*]^{2}[A]-k_{DA}[\R{D}^*]^2[\R{A}^*]+\alpha[A]-\beta [\R{A}^*]\nonumber\\
 & = f([\R{A}^*],[\R{D}^*]), \NT \\
\frac{d[\R{D}^*]}{dt} & = k_{AD}[\R{A}^*]^{2}[D]-k_{DD}[\R{D}^*]^{3} +\gamma [D]-\delta[\R{D}^*]\nonumber\\
 & = g([\R{A}^*],[\R{D}^*]),\NT 
\end{align*}
where $[{\R A}^*],[{\R D}^*]$ are concentration of active species of kinase and phosphatase respectively. The rate constants $k_{ij}$ are caclulated as described in the paper. In our model we assume that the sum of the active and inactive concentration of each species is constant and is set to 1, i.e., $[{\R A}^*]+[{\R A}] = 1$ and $[{\R D}^*]+[{\R D}] = 1$. We find steady state fixed points by setting
\begin{align}
\frac{d[{\R A}^*]}{dt} & = 0, \NT \\
\frac{d[{\R D}^*]}{dt} & = 0. \NT
\end{align}
For a fixed point given by $([\R{A}^*_0],[\R{D}^*_0])$, the Jacobian matrix is

\beq
J=\begin{pmatrix}
f_{[{\R A}^*]}([{\R A}^*_0],[{\R D}^*_0]) & f_{[{\R D}^*]}([{\R A}^*_0],[{\R D}^*_0]) \\
g_{[{\R A}^*]}([{\R A}^*_0],[{\R D}^*_0]) & g_{[{\R D}^*]}([{\R A}^*_0],[{\R D}^*_0])\NT
\end{pmatrix}
\eeq
\\
where,  $f_{[\R{A}^*]},f_{[\R{D}^*]}$ represent partial derivatives with respect to $[\R{A}^*],[\R{D}^*]$ respectively. The eigenvalue of the Jacobian matrix is $\lambda = \mu \pm i\omega$. According to linear stability analysis, $([\R{A}^*_0],[\R{D}^*_0])$ is stable or unstable depending on the real part of $\lambda$, i.e. whether $\mu$ is negative (stable) or positive (unstable). If the system has only one fixed point which is unstable then the system must oscillate. Since we are interested in global oscillators that do not depend on the initial concentration, we look for only those cases where there is only one unstable fixed point.
We find fixed points for a set of dynamical equations using routine fsolve in matlab/octave. Hundreds of trials are performed with different initial values of the dynamic variables ($[A^*]$ and $[D^*]$), and distinct sets of fixed points are sorted out.

\section{Computing PVR and phase shift between the peaks of activators}

To solve the ODEs describing the network of activators and deactivators, we used ode45 function in Matlab. The ODEs are solved for a long enough time interval that the system reaches steady-state oscillations. In Fig. \ref{Oscillations}, we have shown an oscillatory behavior for a 2 activator, 1 deactivator network. In order to compute the peak-to-valley ratio (PVR), we find the peak and the valley (minimum) of the oscillations in steady state and take the ratio of the two. To compute the phase-shift between the activators, we find the time difference between the nearest peak positions of the activators and multiply that by $2\pi/T_{\rm osc}$, where $T_{\rm osc}$ is the period of oscillation.

\renewcommand{\thefigure}{S\arabic{figure}}

\begin{figure}
\includegraphics[width = 8 cm, angle = 0]{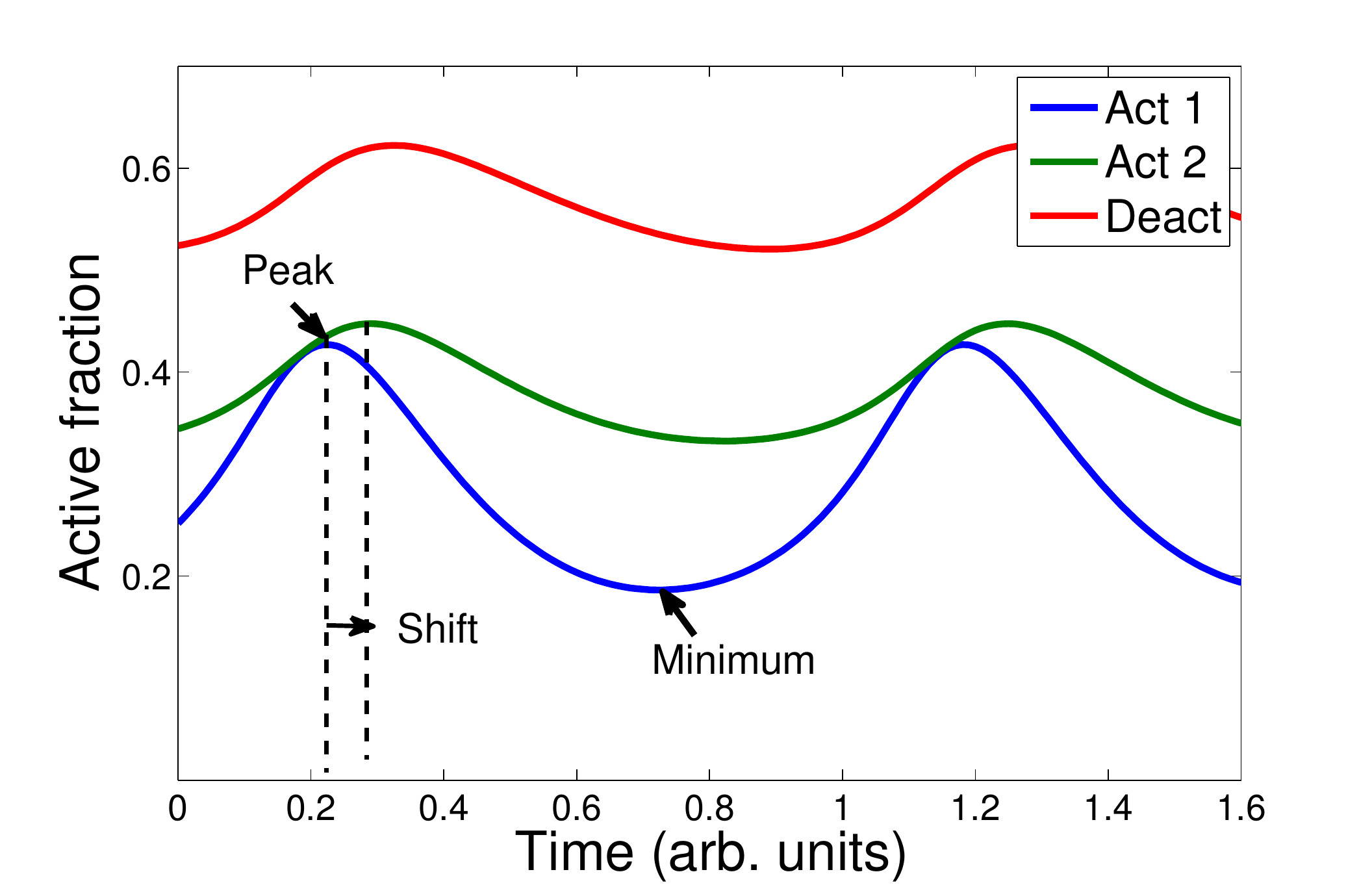}
\caption{Oscillations in 2 activator, 1 deactivator network. The plot shows oscillations of the active fraction of activator 1 (blue), activator 2 (green), and deactivator (red). The dashed vertical lines indicate the peaks of the activator oscillations, and the horizontal arrow indicates the time shift between these peaks.}
\label{Oscillations}
\end{figure}

\section{Principal Component Analysis (PCA)}
In order to examine the correlations between the chemical rate constants and the phases discussed in the main text, we carried out a Principal Component Analysis (PCA) of the rate constants. PCA is usually performed to re-express data in a meaningful basis and to reduce dimensionality (in this case, of the space of rate constants). Our goal is to identify the key rate constants that determine the phase of the system (Phase 1 or 2) for a 3-component system (2 activators, 1 deactivator). For the PCA analysis, we start with a $N_{\scriptscriptstyle T}\times d$ matrix $K$ of rate constants, where each row represents one set of rate constants for a particular mutational step, and each column tracks the time evolution of a rate constant. Here $N_{T}$ is the total number of mutational steps in an evolutionary simulation. For a system of $n$ enzymes, $d=n^2$ is the total number of rate constants, which serves as the dimension for PCA analysis; for our 3-component system the dimension is $d=9$. We follow the following standard steps for PCA:
\begin{itemize}
\item
Subtract the mean of each dimension (i.e. each rate constant) from the corresponding column of $K$ to construct a matrix $X$.
\item
Calculate the covariance matrix $C$ of $X$ : $C=\frac{1}{N_{\scriptscriptstyle T}-1}X^TX$.
\item
Find the eigenvalues and corresponding eigenvectors of the covariance matrix $C$.
\item
To reduce the dimensionality, consider only the largest $m$ eigenvalues and form an orthonormal feature matrix from the corresponding eigenvectors $M = [eig_1$  $eig_2 ...$ $eig_m]$.
\item
Finally multiply the mean-subtracted data $X$ by the feature vector $M$ to obtain a projected data set of reduced dimension $N_{T}\times m$. i.e. $ \R{Final Data} = XM$.
\end{itemize}

The eigenvector with the largest eigenvalue corresponds to the principal-component direction along which the data has the largest variance. For the data obtained from a very long evolutionary trajectory, we find that the eigenvalues of the covariance matrix are, in ascending order, [$0.002$ $0.019$ $0.027$ $0.2$ $0.229$ $0.557$ $0.811$ $1.688$ $8.214$] and the eigenvectors (columns) corresponding to each eigenvalue are\\
\\
\begin{tabular}{c | c c c c c c c c c }
	&eig1	&eig2	&eig3	&eig4	&eig5	&eig6	&eig7	&eig8	&eig9\\
\hline
$k_{11}$	&0.015 &   0.117   & 0.052 &   0.399   & 0.485 &  -0.244  & -0.358 &   0.355  & -0.525\\
$k_{12}$	&0.041  &  0.034   & 0.036  & -0.372   & 0.387  &  0.802  & -0.068 &   0.232  & -0.081\\
$k_{13}$	&-0.149  & -0.968  & -0.135 &   0.039  &  0.126 &  -0.006 &  -0.032 &   0.042 &  -0.034\\
$k_{21}$	&0.029   & 0.022   & 0.031  &  0.288   & 0.324  & -0.004  &  0.875  &  0.199  &  0.067\\
$k_{22}$	&0.027   & 0.009   & 0.164  &-0.396    & 0.4  & -0.429  & -0.142  &  0.311  &  0.595\\
$k_{23}$	&-0.161  &  0.171  & -0.954  & -0.097  &  0.126 &  -0.071 &   0.006 &   0.049 &   0.043\\
$k_{31}$	&-0.05  & -0.036  &  0.019  & -0.457  & -0.434 &  -0.193 &   0.21 &   0.589 &  -0.412\\
$k_{32}$	&-0.061  &  0.006  & -0.058  &  0.492 &  -0.356 &   0.267 &  -0.189 &   0.575 &   0.432\\
$k_{33}$	&0.971   &-0.126  & -0.189   & 0.009 &  -0.049  & -0.024 &  -0.019  &  0.051  &  0.001\\
\end{tabular}
\\
\\
Note that the rows here correspond to the rate constants $k_{11}$, $k_{12}$, $k_{13}$, $k_{21}$, $k_{22}$, $k_{23}$, $k_{31}$, $k_{32}$, $k_{33}$, respectively. To produce Fig.~3E of the main text, we employed the eigenvector PC1 corresponding to the largest eigenvalue (the last column of the above matrix).
\section{Evidence for geometric bottleneck}

Here we review briefly our evidence for a geometric bottleneck between the two phases for 3-component oscillators (2 activators and 1 deactivator). The idea was initially introduced in the main text to explain the observed long-term memory related to the order of magnitude difference between flipping of essentiality and flipping of phase. This idea of a bottleneck found support in the exponential distribution of duration in a single phase (between two phase flips). As noted in the manuscript, we checked this hypothesis of the bottleneck by studying the neighborhood of states in Phase 1 and Phase 2. For example, for only 3\% of the states in Phase 1 where both activators are essential does the Hamming distance 1 neighborhood contain one or more states where Activator 1 is inessential. The relative rarity of such states (which can be considered as boundary states) is consistent with our hypothesis that in sequence space the two phases touch at a relatively small number of boundary points 

\begin{figure}
\includegraphics[width = 15 cm, angle = 0]{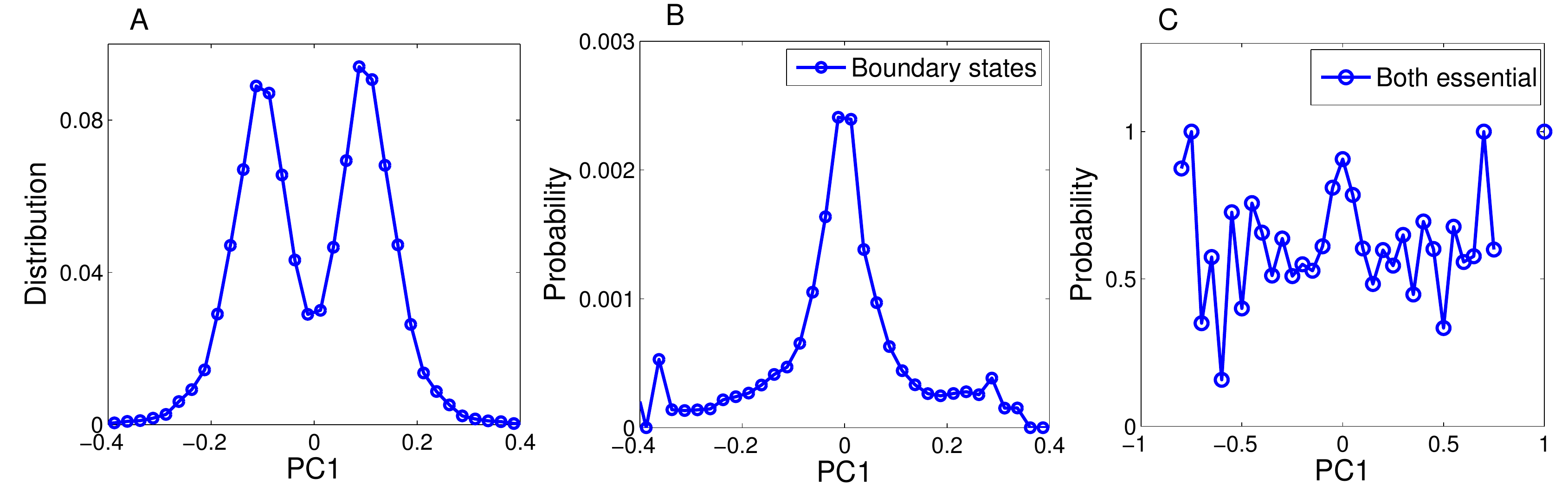}
\caption{Distribution of states of a 2 activator, 1 deactivator system. (A) Distribution of PC1. (B) Probability of boundary states between the two phases. (C) Probability of states where both activators are essential.}
\label{Geom_1}
\end{figure}

  Due to the difficulty of visualizing a very high-dimensional sequence space, it is also helpful to study the distribution of points belonging to the two phases in the space of the chemical rate constants. As noted in the manuscript, one direction in this space, designated as PC1 (principal component 1), was relatively effective in discriminating between the phases (though due to some overlap, the discrimination was not complete, as can be seen in Fig.~2E in the main text). Thus in order to characterize the geometric bottleneck, we plot the distribution of states in the PC1 direction (Fig.~\ref{Geom_1}A). The distribution is bimodal with the two peaks representing the two phases (with the left peak corresponding to Phase 1 and the right to Phase 2) well separated out. In Fig.~\ref{Geom_1}B, we plot the probability that a state is a boundary state (as identified in the main text) along the PC1 direction and, as expected, find a single narrow peak at the center. We note that even near this peak the probability of being in a boundary state is very small ($\sim 10^{-3}$) which also suggests that there are relatively very few boundary states, supporting the geometrical bottleneck hypothesis. In contrast, the states where both the activators are essential are uniformly distributed along PC1 (Fig.~\ref{Geom_1}C) and such states represent $\sim66\%$ of all oscillatory states.  
  
As an aside, we also checked if there is any significant difference in the probability that a mutation will fail (that is, lead to a non-oscillatory state) for boundary points versus interior points or, alternatively, as a function of PC1, since a higher value of the failure probability for boundary points could potentially further enhance memory by penalizing genotypes near the transition between the two phases. However, perhaps surprisingly, we find no significant difference in the frequency with which mutations fail for boundary points versus points interior to the two phases.
\section{Toy Model}

Due to the high dimensionality of the sequence space and the complex nature of the relationship between the model parameters and observed behavior of flipping of essentiality and phases, we found it helpful to construct a simplified toy model that can reproduce crucial aspects of the observed behavior.
Specifically, we introduce a toy model consisting of two activators, Activator 1 (Act1) and Activator 2 (Act2). The essentiality of each activator is determined by a binary string (consisting of 0s and 1s) of even length $L$. In our toy  model, Act1 (Act2) will be essential if the number of 1s in the first half (second half) of the binary string is greater than or equal a threshold $N_c$. We evolve the string by randomly flipping bits, and requiring that at least one activator is always essential. We also constrain the total sum of 1s in the whole string to be equal to or less than a cutoff, $M_c$, which is greater than or equal to $2N_c$ so that both activators can in principle be essential at the same time. Setting the difference between $M_c$ and $2N_c$ to be small leads to relatively few states where both activators are essential, and thus creates a ``bottleneck''  between domains where each activator is essential.

\begin{figure}
\includegraphics[width = 16 cm, angle = 0]{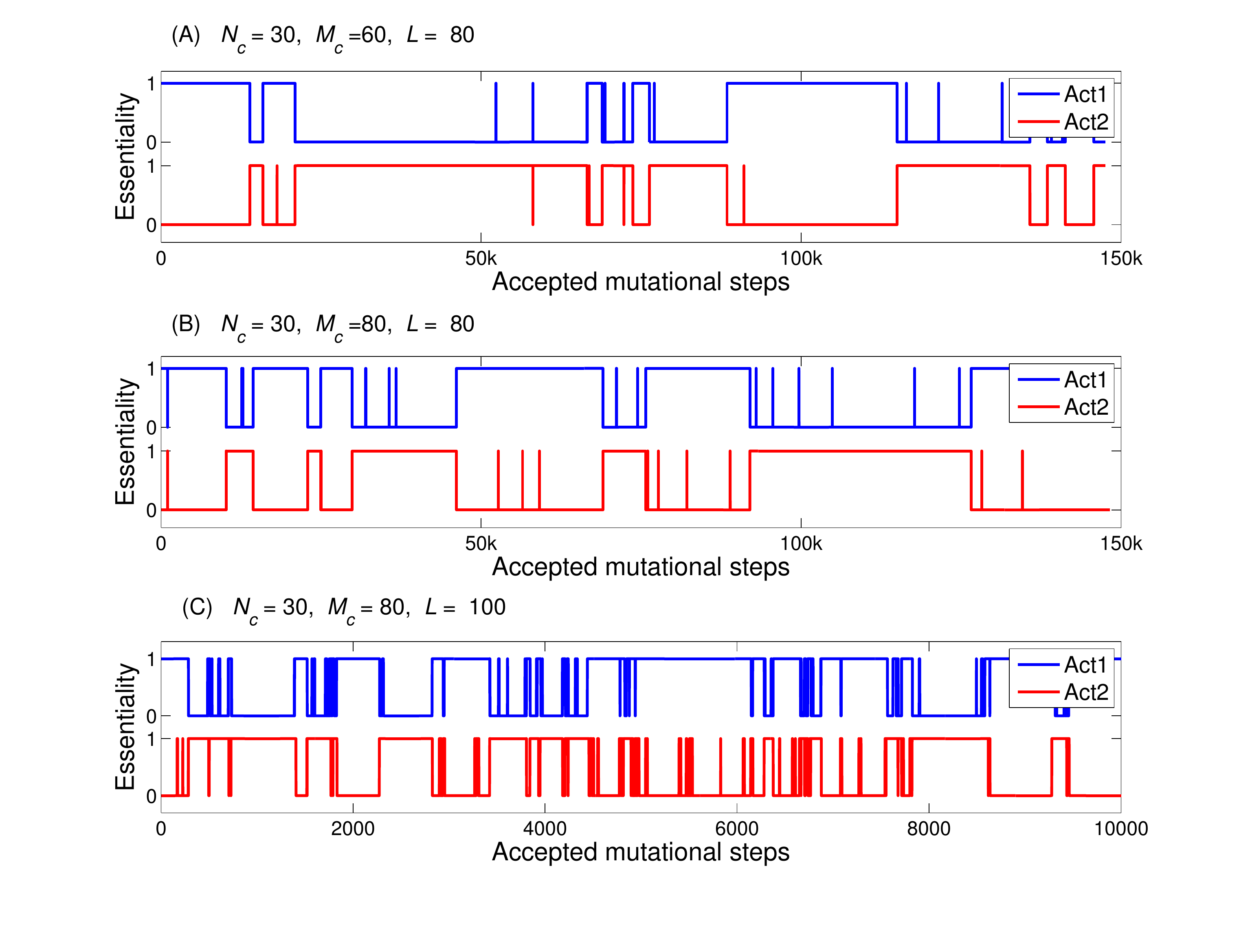}
\caption{Temporal evolution of essentiality for two activators in the toy model. A value of +1 indicates essential and 0 indicates inessential. We start with string such that Activator 1 is essential (first half of the string has at leas $N_c$ 1s) to begin with and evolve with the constraints that one of the activators is always essential and the total number of 1s is less than or equal to $M_c$.}
\label{Toy-1}
\end{figure}

\begin{figure}
\includegraphics[width = 14 cm, angle = 0]{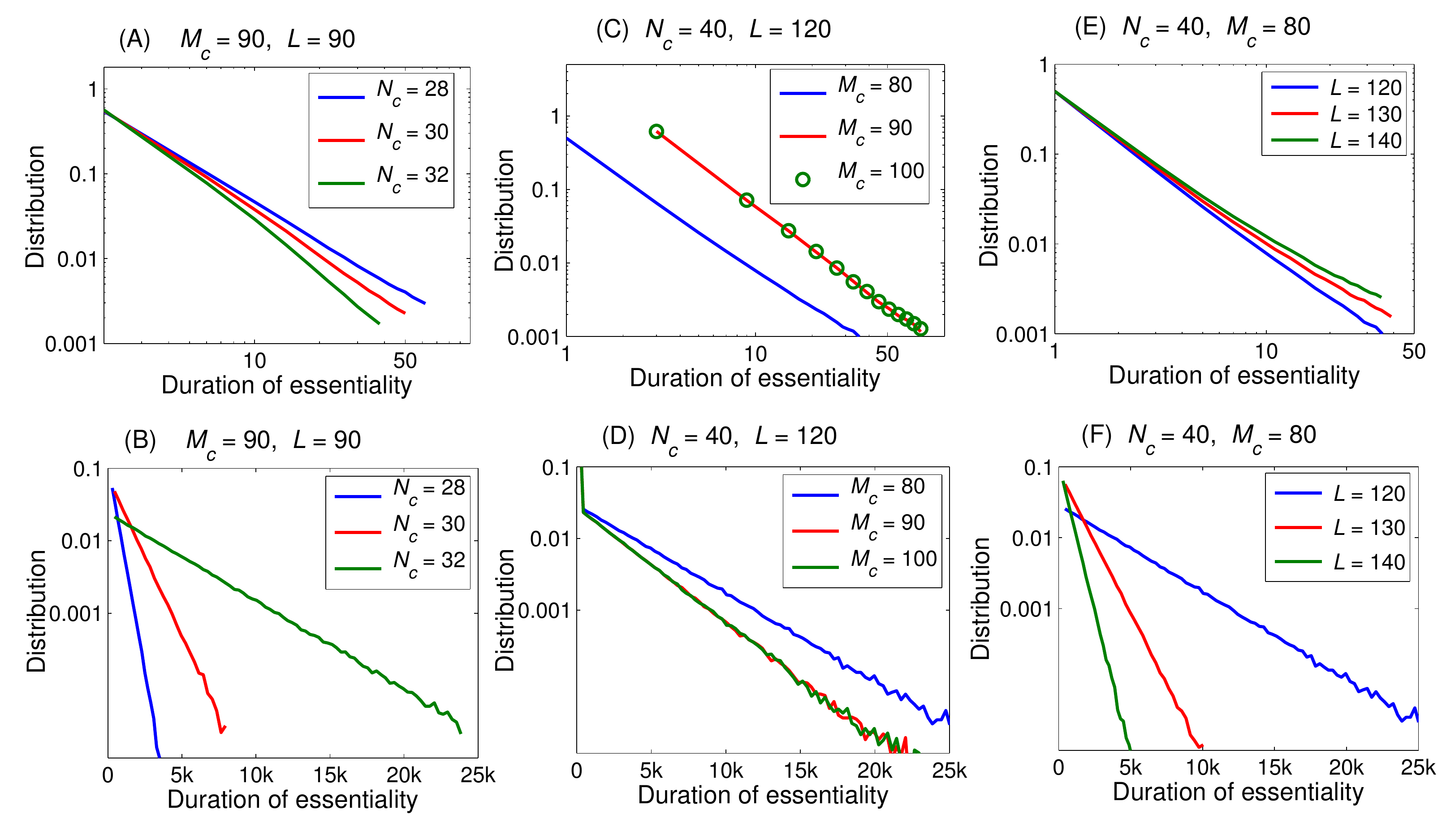}
\caption{Distribution of the number of accepted mutational steps where an activator is essential for the whole duration, (A,C,E) on a log-log scale showing a power-law behavior  for short durations, and (B,D,F) on a semi-log scale showing exponential decay over longer durations on semi-log scale.}
\label{Toy-3}
\end{figure}

\begin{figure}
\includegraphics[width = 14 cm, angle = 0]{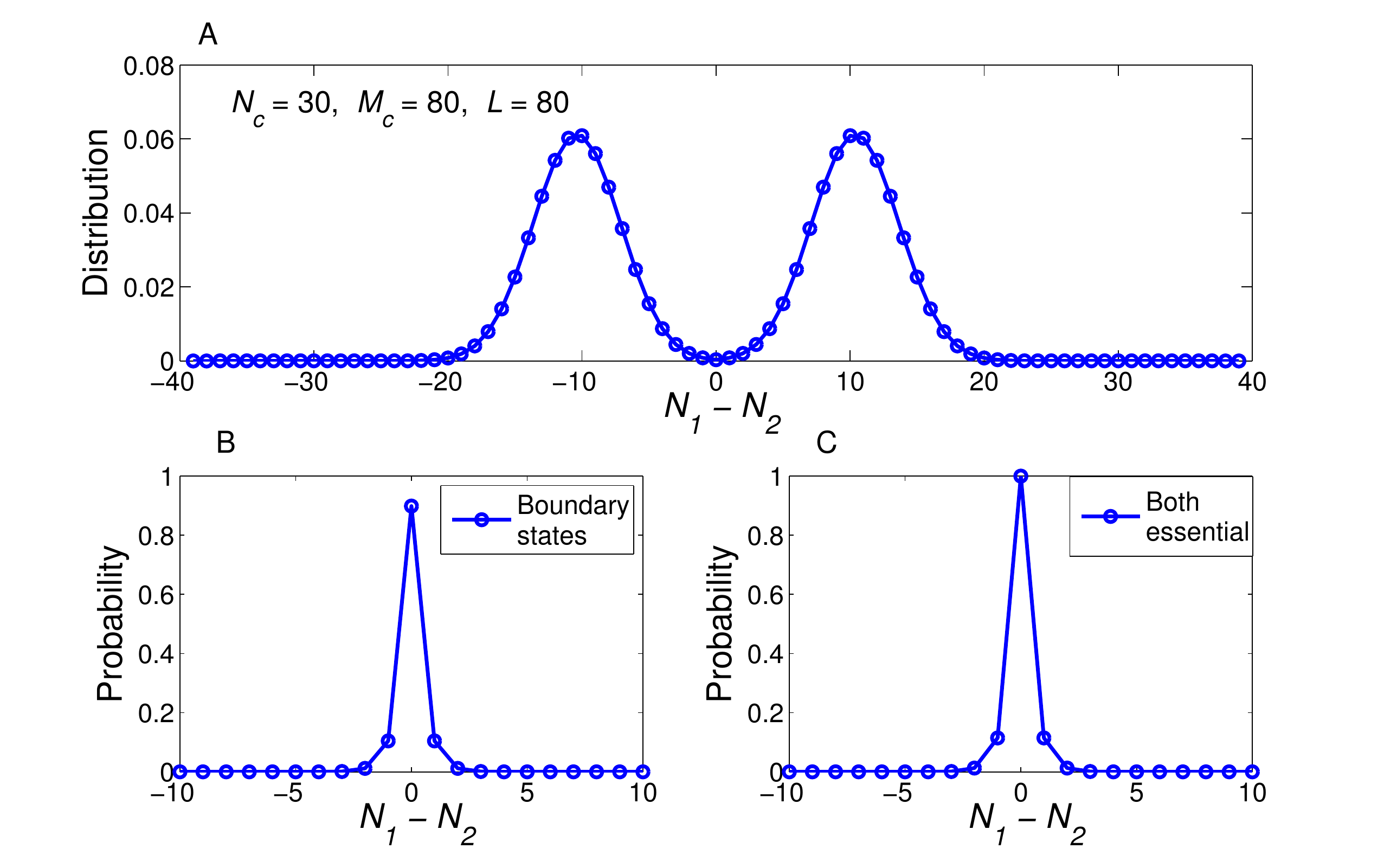}
\caption{(A) Distribution of $\Delta N = N_1-N_2$ in the toy model for $N_c = 30,~M_c = 80$, and $L = 80$. (B) Probability of boundary states. (C) Probability of states where both activators are essential.}
\label{Toy-4}
\end{figure}

The toy model exhibits behavior that is similar in some crucial respects to the evolutionary model presented in the main text, but also displays some important differences. For our toy model, we start the system with binary string such that Act1 is essential and Act2 inessential, and let the system evolve by introducing random point mutations. We depict the time evolution of essentiality of activators in Fig.~\ref{Toy-1} for different values of the parameters ($N_c$, $M_c$, and $L$). As with our evolutionary model, we find flips in essentiality of both Act1 and Act2, with a power-law distribution of the number of mutational steps that an activator remains essential ($P(T) \propto T^{- \alpha}$) for short durations, and exponential decay ( $P(T) \propto e^{- T/\tau_0}$ ) for longer durations. For the power-law distribution, the exponent $\alpha$ depends on the model parameters, $N_c$, $M_c$, and $L$, as follows. For a fixed value of $M_c$ and $L$, $\alpha$ increases with the threshold $N_c$ (Fig.~\ref{Toy-3} A). The exponent does not vary with $M_c$ for fixed $N_c$ and $L$ except when $M_c$ is equal to $2N_c$ (Fig.~\ref{Toy-3}~C). For a fixed value of $N_c$ and $M_c$ the exponent decreases with the length of the sequence (Fig.~\ref{Toy-3}~E). For the exponential decay at longer durations, $P(T) \propto e^{- T/\tau_0}$, the dependence of the constant $\tau_{0}$ on the model parameters resembles that of the exponent $\alpha$ (Fig.~\ref{Toy-3}~B,D,F).
To understand the behavior generated by our toy model, it is worth noting that the sequence space can be characterized by two variables, $N_1$ and $N_2$, where $N_1$ ($N_2$) is the number of 1s in the first half (second half) of the string. We note that for any allowed sequence with $N_1 > N_2~(N_2 > N_1)$, Act1 (Act2) has to be essential. Thus we can characterize these points as Phase 1 (Phase 2). In the toy model $\Delta N=N_1-N_2$ plays a role similar to principal component (PC1 in the full model). We choose our parameters such that the distribution of $\Delta N$ is bimodal (Fig.~\ref{Toy-4}~A), similar to the distribution of PC1. Note that the distribution of phases yields similar decay time as the decay of essentiality. 

\begin{figure}
\includegraphics[width = 14 cm, angle = 0]{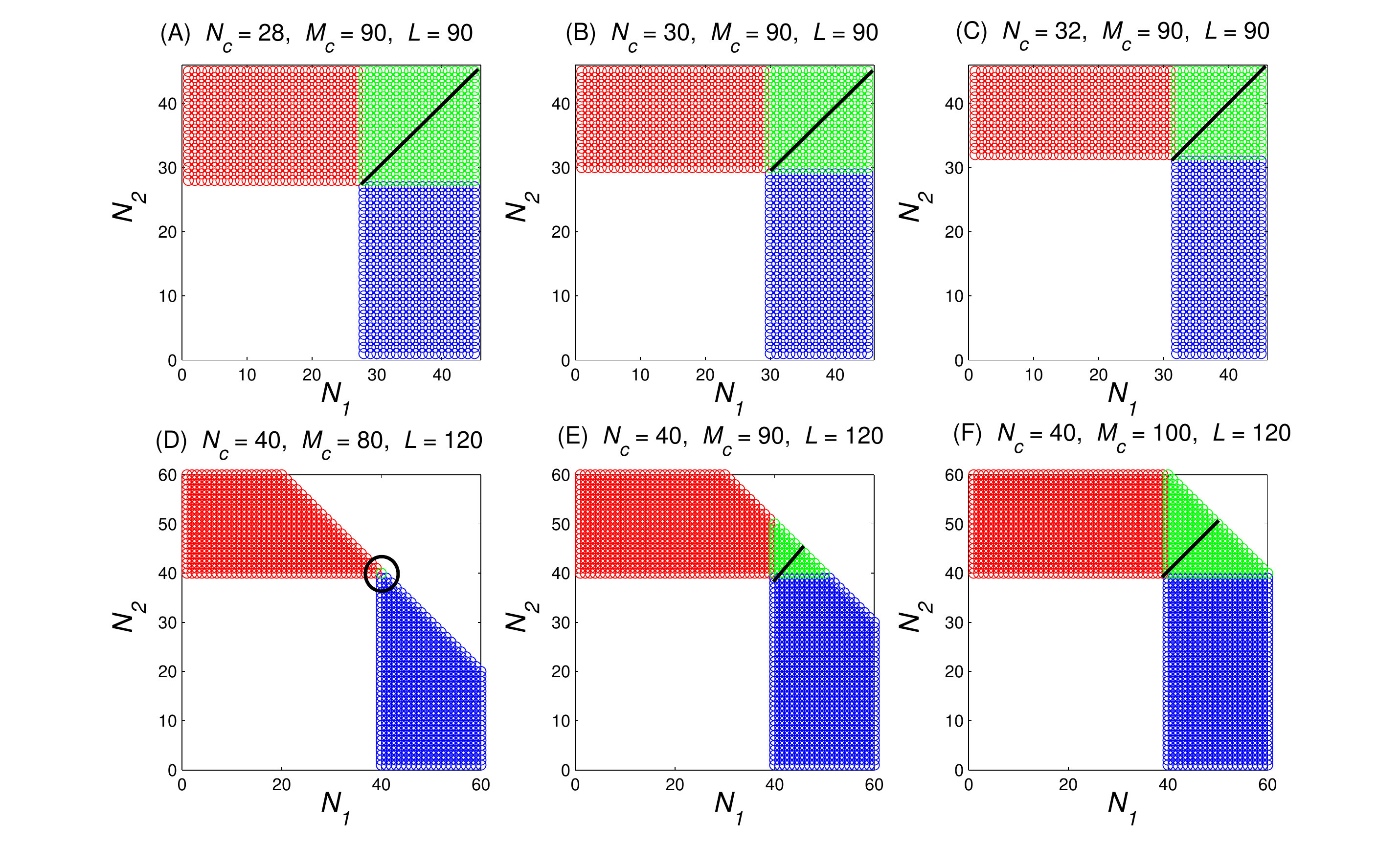}
\caption{Essentiality of activators in $N_1-N_2$ space. Activator 1 essential (red), Activator 2 essential (blue), both essential (green). Solid black line represents the boundary states separating the two phases.}
\label{Toy-2}
\end{figure}

\begin{figure}
\includegraphics[width = 16 cm, angle = 0]{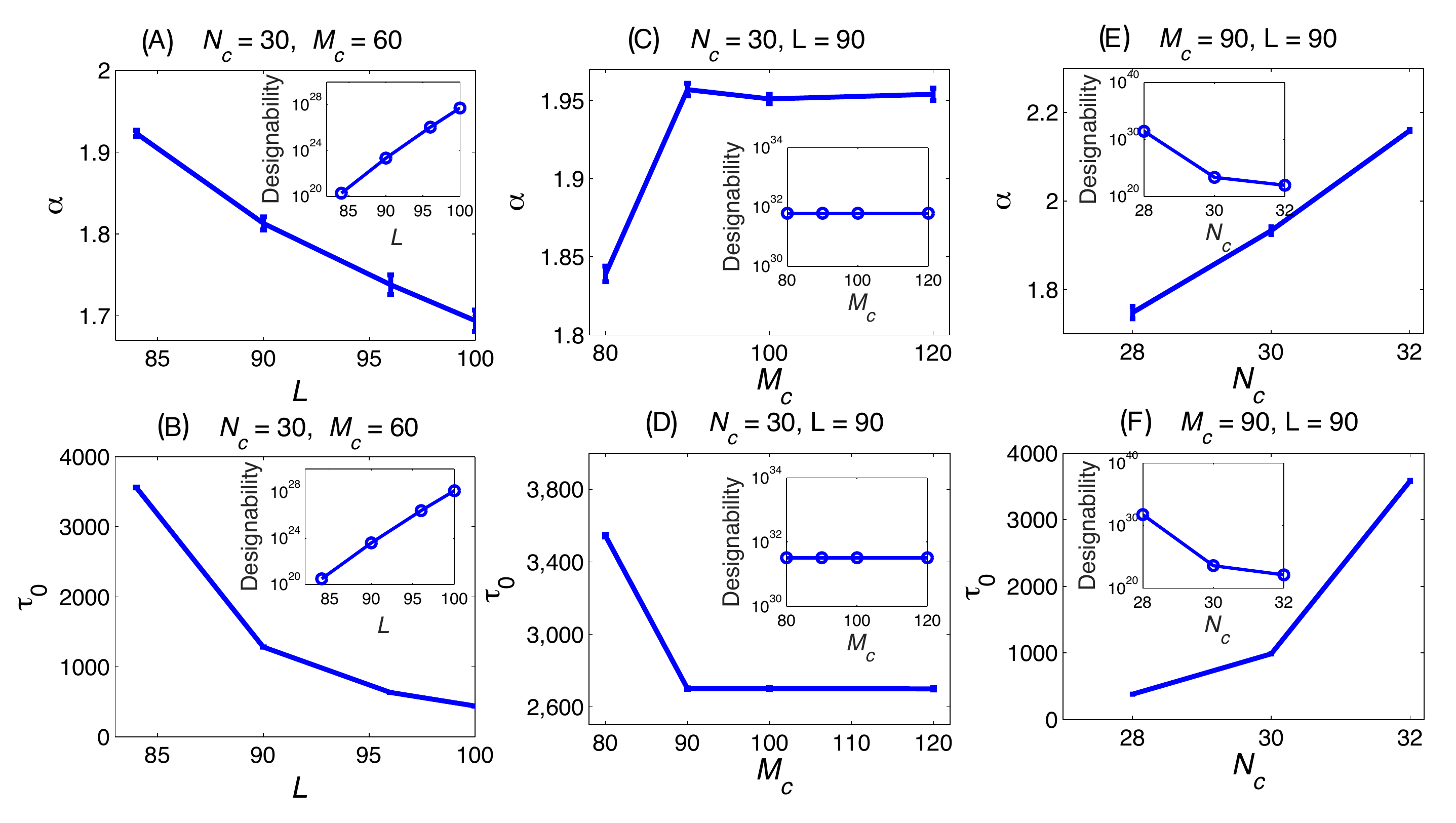}
\caption{(A,C,E) Variation of power-law exponent $\alpha$ with toy model parameters. Inset: Constrained designability versus model parameters. (B,D,F) Variation of exponential decay time $\tau_0$ with toy model parameters. Inset: Designability of boundary region versus model parameters.}
\label{Toy-con}
\end{figure}

In order to understand how the exponent and decay time depends on the model parameters ($N_c$, $M_c$, and $L$), we depict, in Fig.~\ref{Toy-2},  the essentiality of each activator in $N_1$-$N_2$ space. The green region corresponds to points where both activators are essential, and, for our toy model, it also corresponds to the boundary region between the two phases. For any allowed point in $N_1$-$N_2$ space, 
 the designability (the number of distinct allowed sequences) is given by $^{L/2}\!\!~C_{N_1} \times ^{L/2}\!\!~C_{N_1}$. For given values of  $N_c$, $M_c$, and $L$, we can calculate the total designability associated with the boundary region (total designability equals the sum of the designability of all allowed points in the boundary region of $N_1$-$N_2$ space).  As expected, we find that that dependence of $\tau_{0}$ on any one of the model parameters (with the other two parameters held fixed) correlates with total designability. A lower designability (which could correspond to a more pronounced bottleneck between the phases) indicates longer phase duration and, correspondingly, larger value of $\tau_0$. For the power-law exponent, we similarly expect the exponent to be related to the boundary of the green region in Fig. ~\ref{Toy-2}. A measure of the boundary is the constrained sum of the designabilities of only those states in the green region where either $N_{1}$ or $N_{2}$ equals $N_{c}$, and we find that $\alpha$ indeed correlates with this constrained designability. To be more precise, if any of the model parameters is varied holding the other two fixed, higher constrained designability indicates lower value of $\alpha$ and vice-versa. In Fig.~\ref{Toy-con}, we have shown the dependence of exponent $\alpha$ and decay time $\tau_0$ on model parameters along with the relevant designabilities.

Despite the similarities highlighted in the previous paragraphs, the behavior of the toy model also differs in some crucial respects from the behavior exhibited by the full model in the main text. For example, in the toy model, we typically do not find long durations where one of the activator remains essential while the other flips essentiality. Rather, the mean phase duration and mean duration of essentiality are similar (differing by a factor of 2-3, in contrast to the full model where they differ by an order of magnitude). Moreover, if the distribution of $\Delta N$ is chosen to be bimodal, we find that typically the fraction of time where both activators are essential is very small ($\sim1\%$), whereas in the full model this fraction is high ($\sim66\%$). In the toy model, the fraction of time both activators are essential is determined by the ratio between $N_c$ and $L/4$. When this ratio is smaller than 1, it is rare for both activators to be essential. As this ratio approaches 1, it becomes increasingly common for both activators to be essential, and if the ratio becomes greater than 1, then both will be essential most of the time. However, when the ratio is greater than 1 the distribution of $\Delta N$ is no longer bimodal. Moreover, 
unlike the results from the full model, in the toy model, the region in $N_1$-$N_2$ space where both activators are essential forms a compact domain (shown in green in Fig.~\ref{Toy-2}) and lies at the boundary between the two phases. The difference is highlighted by the density of states in the PC1/$\Delta N( = N_1-N_2)$  direction. In the full model, the distribution of states where both activators are essential is almost uniform in the PC1 direction (Fig.~\ref{Geom_1}~A), quite different from the distribution of boundary points which is peaked around the center of the PC1 axis (Fig.~\ref{Geom_1}~B). This contrasts with the toy model, where the distribution of states where both activators are essential as well as the distribution of boundary states are peaked around the center of $\Delta N( = N_1-N_2)$ (Fig.~\ref{Toy-4}~B, C).

In summary, we have shown that the toy model captures some of the features of the full activator-deactivator oscillator model, e.g., the presence of mixed power-law and exponential distributions of flipping of essentiality. However, unlike the full model, the toy model does not exhibit long-term evolutionary memory; the difference in behavior can be related to the distribution of states in sequence space where both activators are essential.

\clearpage
\section{Starting sequence for 2 activator, 1 deactivator system}
The starting sequence for the network shown in green in Fig.~2 is\\
\\
\begin{tabular}{l*{23}{c}r}
1 & 1 & 1 & 1 & 1 & 1 & 1 & 1 & 1 & 1 & 1 & 1 & 1 & 1 & 0 & 0 & 0 & 0 & 0 & 0 & 0 & 0 & 0 & 0 & 0\\
1 & 1 & 1 & 1 & 1 & 1 & 1 & 1 & 1 & 1 & 1 & 1 & 1 & 1 & 1 & 1 & 1 & 1 & 1 & 1 & 0 & 0 & 0 & 0 & 0\\
0 & 0 & 0 & 0 & 0 & 0 & 0 & 0 & 0 & 0 & 0 & 0 & 0 & 0 & 0 & 0 & 0 & 0 & 0 & 0 & 0 & 0 & 0 & 0 & 0\\
0 & 0 & 0 & 0 & 0 & 0 & 0 & 0 & 0 & 0 & 0 & 0 & 0 & 0 & 0 & 0 & 0 & 0 & 0 & 0 & 0 & 0 & 0 & 0 & 0\\
0 & 0 & 0 & 0 & 0 & 0 & 0 & 0 & 1 & 1 & 1 & 1 & 1 & 1 & 1 & 1 & 1 & 1 & 1 & 1 & 0 & 0 & 0 & 0 & 0\\
1 & 1 & 1 & 1 & 1 & 1 & 1 & 1 & 1 & 1 & 1 & 1 & 0 & 0 & 0 & 0 & 0 & 0 & 0 & 0 & 0 & 0 & 0 & 0 & 0\\
\end{tabular}
\\
\\
where rows $1,3,5$ are the out-faces of Activator 1, Activator 2, and the deactivator, and $2,4,6$ are the in-faces for the same. The zeros in row 3 and 4 indicates that Activator 2 is minimally interacting with Activator 1 and the deactivator. The interaction energies corresponding to this sequence for $\epsilon = 0.2$ are [$E_{11}$ $E_{12}$ $E_{13}$ $E_{21}$ $E_{22}$ $E_{23}$ $E_{31}$ $E_{32}$ $E_{33}$] = [ 2.8 0.0 2.4 0.0 0.0 0.0 2.4 0.0 0.8 ]. The subscript $1,2,3$ in the energies denotes Activator 1, Activator 2, and the deactivator.\\

The starting sequence for the network shown in blue in Fig.~2 is\\
\\
\begin{tabular}{l*{23}{c}r}
1& 1& 1& 1& 1 & 1 & 1 & 1 & 1 & 1 & 1 & 1 & 1 & 1 & 0 & 0 & 0 & 0 & 0 & 0 & 0 & 0 & 0 & 0 & 0\\
1& 1& 1& 1& 1 & 1 & 1 & 1 & 1 & 1 & 1 & 1 & 1 & 1 & 1 & 1 & 1 & 1 & 0 & 0 & 0 & 0 & 0 & 0 & 0\\
0& 0& 0& 0& 0 & 0 & 0 & 0 & 0 & 0 & 0 & 0 & 0 & 0 & 0 & 0 & 0 & 0 & 0 & 0 & 0 & 0 & 0 & 0 & 0\\
0& 0& 0& 0& 0 & 0 & 0 & 0 & 0 & 0 & 0 & 0 & 0 & 0 & 0 & 0 & 0 & 0 & 0 & 0 & 0 & 0 & 0 & 0 & 0\\
0& 0& 0& 0& 0 & 1 & 1 & 1 & 1 & 1 & 1 & 1 & 1 & 1 & 1 & 1 & 1 & 1 & 0 & 0 & 0 & 0 & 0 & 0 & 0\\
1& 1& 1& 1& 1 & 1 & 0 & 0 & 0 & 0 & 0 & 0 & 0 & 0 & 0 & 0 & 0 & 0 & 0 & 0 & 0 & 0 & 0 & 0 & 0\\
\end{tabular}
\\
\\
and the corresponding interaction energies are [$E_{11}$ $E_{12}$ $E_{13}$ $E_{21}$ $E_{22}$ $E_{23}$ $E_{31}$ $E_{32}$ $E_{33}$] = [ 2.8 0.0 2.6 0.0 0.0 0.0 1.2 0.0 0.2 ].\\
We studied the time evolution of the essentiality of each activator for the above two starting sequences. We found by averaging over multiple long runs that on average $61.6 \%$ of the time both activators are essential. We also found that states where both are inessential are very rare, approximately $0.001\%$ of the total number of oscillatory states. For $38.4\%$ of the time only one of the activators is essential.

\section{Correlation between relative PVR, sign of phase shift, essentiality, and rate constants}

Why are relative peak-to-valley ratio (PVR), sign of phase shift, essentiality, and relative rates of activator auto-activation/deactivation correlated? We offer here a qualitative argument for a 3-component system (2 activators and 1 deactivator). If Activator 1 is leading in phase, it likely has a stronger self-activation than the second activator and thus its activated level $\R{A}_1^{*}$ starts increasing once the level of activated deactivator $(\R{D}^{*})$ is sufficiently low. Once the level of Activator 1 becomes sufficiently high, it starts activating both Activator 2 and the deactivator, whose levels both start rising. Once the $\R{D}^{*}$ level builds up it starts deactivating both the activators, whose active levels then start to drop. However, since $\R{A}_1^{*}$ had started rising first, the phase shift will be positive and, furthermore, $\R{A}_1^{*}$ will also have reached a relatively higher level than $\R{A}_2^{*}$  before both levels start to drop, so that Activator 1  exhibits a higher PVR. Moreover in this scenario, Activator 1 might be expected to play a more important role for the oscillations since it is the auto-activation of Activator 1 that drives both its level and the activated level of Activator 2 to rise, and is thus more likely to play an essential role in the oscillations. It is important to note here that this argument is only qualitative and does not necessarily apply to all oscillatory states. 

\section{Exponential versus power-law distribution of duration times}
Consider a system that can be in two phases: Phase 1 and Phase 2. Furthermore, consider the case where the system enters one of the two phases, say Phase 1, at time $t=0$. We seek to find an expression for the probability distribution of the time of duration in Phase 1 before the system switches to Phase 2. If $P_{0}(t)$ is the probability that system still persists in Phase 1 at time $t$ without having switched to Phase 2, and $P_{T}(t)$ is the probability density of duration time $t$, then $dP_{0}/dt = -P_{T}(t)$. If the switching rate from Phase 1 to Phase 2 is a constant $k$, independent of time elapsed since entry into Phase 1, then
\begin{equation}
 \frac{d P_{0}}{d t} = - k P_{0}, \NT 
 \end{equation}
 implying
 \begin{equation} 
 P_{0} (t) = e^{- k t}. \NT 
 \end{equation}
 In this case, the probability density $P_{T}(t)$ is also exponential, $P_{T} \propto e^{- k t}$. We expect this to be the case for an entropic barrier separating the two phases. 

In contrast, if the escape rate $k$ itself depends on time $t$ elapsed from the moment of entry into Phase 1, in particular, if it is of the form $k = \alpha /(\tau_{0} + t)$, where $\tau_{0}$ and $\alpha$ are constants, (we might expect such a form for an extended boundary between the two phases, since the longer the system has spent in Phase 1, the deeper, that is further from the boundary, it is likely to be in the phase; also $\tau_{0}$ is roughly of the order of the timescale for a single mutational step), then
\begin{equation}
 \frac{d P_{0}}{d t} = - \frac{\alpha}{\tau_{0} + t} P_{0}, \NT 
 \end{equation}   
 implying 
 \begin{equation}
 P_{0}(t) = \left(\frac{\tau_{0} + t}{\tau_{0}} \right)^{-\alpha}. \NT 
 \end{equation}
 In this case, for $t > \tau_{0}$, we obtain a power-law distribution $P_{T}(t) \sim t^{-a}$, where $a = \alpha - 1$.

\section{Generalization to 5 activator, 5 deactivator system}
\begin{figure}
\includegraphics[width = 12 cm, angle = 0]{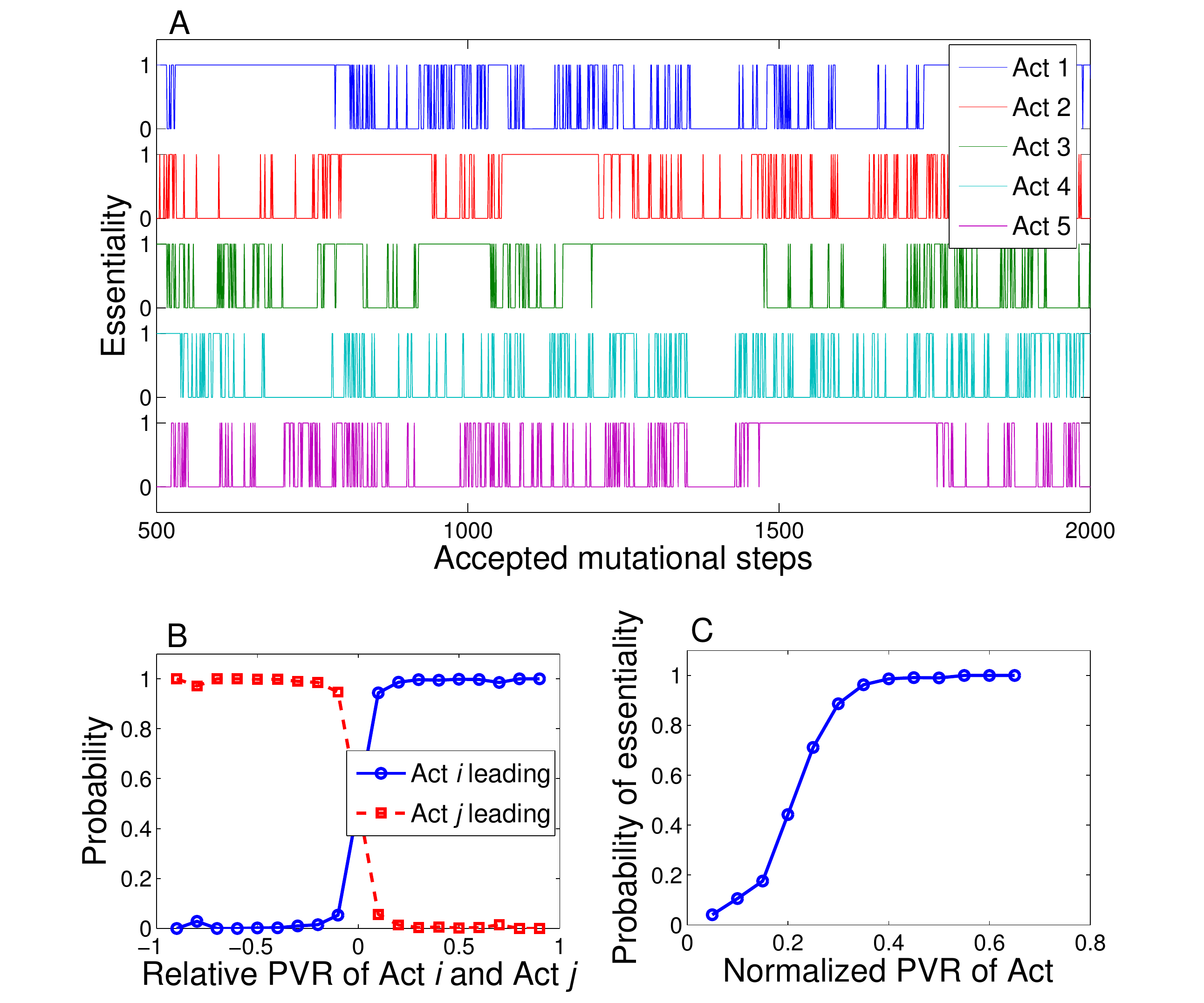}
\caption{(A) Temporal evolution of essentiality of the five activators for a 10-component system of 5 activators and 5 deactivators. On the $y$-axis, +1 indicates that the activator is essential and 0 indicates that it is not essential. We notice durations where an activator remains essential and durations where it continues to flip between being essential and inessential.  (B) For any pair of activators, the probability Activator $i$ leads Activator $j$ as a function of their relative PVR. (C) Probability of an activator, say Activator $i$, being essential as a function of its normalized PVR defined as PVR A$_i$/($\sum_{j = 1}^{5}$ PVR A$_j$).}
\label{TenComp}
\end{figure}

In order to check the applicability of our main results regarding essentiality of activators, relative phase, and  PVR  for more complex networks, we studied a network of $5$ activators and $5$ deactivators. To simply generate a 5 activator, 5 deactivator oscillator, we started with a 2-component (1 activator, 1 deactivator) oscillator and divided both the activator and deactivator into 5 identical copies, each with the same set of sequences as its parent. We set the concentrations of each of the 5 new activators and 5 new deactivators to be 1/5 of its parent, so that initially the dynamics of the system was identical to the starting 2-component oscillator. The system was evolved for  $\sim$3000 accepted mutational steps such that the sequences for the activators and deactivators became quite different. The system was further evolved for $25,000$ mutational steps from which we obtained results for essentiality, relative phase, and PVR. During network evolution, mutations were accepted if the system continued to oscillate for a given initial concentration ($0.5$ for each active fraction). Specifically, for each proposed mutation we solved the dynamical equations and accepted only those mutations for which the amplitude of oscillation of the active fraction of at least one of the components remained above a cutoff ($0.001$ in this case) for 400 units of time. To test the essentiality of each activator we removed that component and checked if the system continued to oscillate. The plot of essentiality of the 5 activators is shown in Fig.~S1A. We observed durations where a given activator remains essential and durations where it continues to flip between essential and inessential.  This behavior is very similar to what we observed for a 4-component oscillator (Fig.~5). For a given pair of activators we calculated the PVR and also tracked which activator was leading in phase in the oscillations. We found that similar to 4-component oscillators, for any pair of activators, the probability of one activator leading the other is higher if the relative PVR is higher (Fig.~S1B), and the probability that an activator is essential is higher when its normalized PVR is higher (Fig.~\ref{TenComp}C).

\section{Robustness of the model}
\begin{figure}
\includegraphics[width = 8 cm, angle = 0]{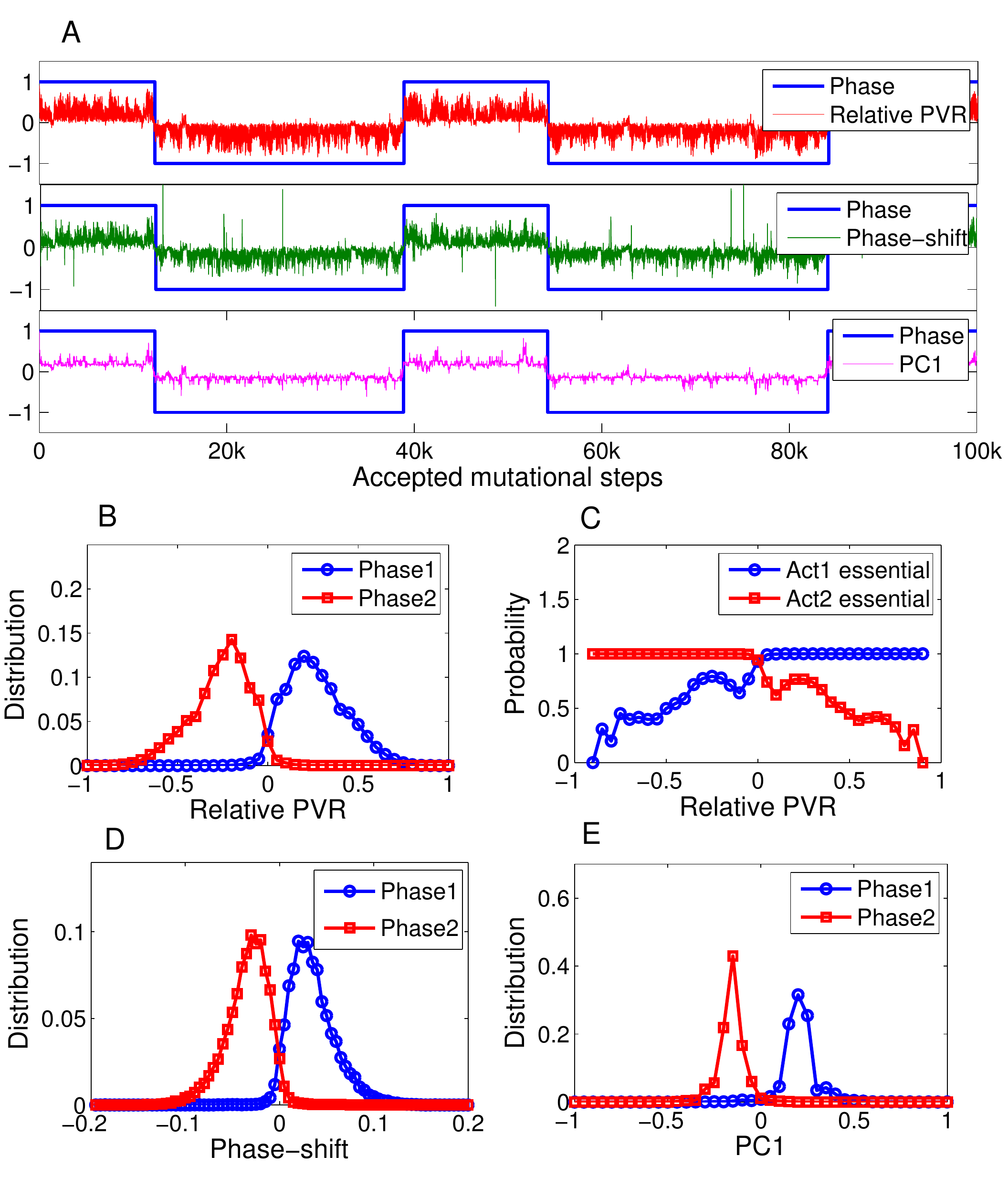}
\caption{Temporal evolution of phases in 2 activator, 1 deactivator system for $k_0 = 5625$, $\epsilon = 0.2$, and $\alpha = \alpha^{'} = 1$. (A) Depiction of the temporal evolution where a value of +1 indicates Phase 1 and -1 indicates Phase 2. Along with the phase, the three panels show (i) normalized relative PVR of the two activators (red), (ii) phase-shift between their oscillatory peaks (green), and (iii) projected component of the chemical rates on the principal eigenvector from PCA analysis (magenta). (B) Distributions of relative PVR of the two activators in Phase 1 and in Phase 2. (C) Probability that each activator is essential as a function of its relative PVR. (D) Distribution of phase-shifts between active fraction peaks of the two activators in Phase 1 and Phase 2. (E) Distribution of projected rate constants on the principal eigenvector, obtained from PCA analysis, in Phase 1 and Phase 2.}
\label{Robust1}
\end{figure}

\begin{figure}
\includegraphics[width = 14 cm, angle = 0]{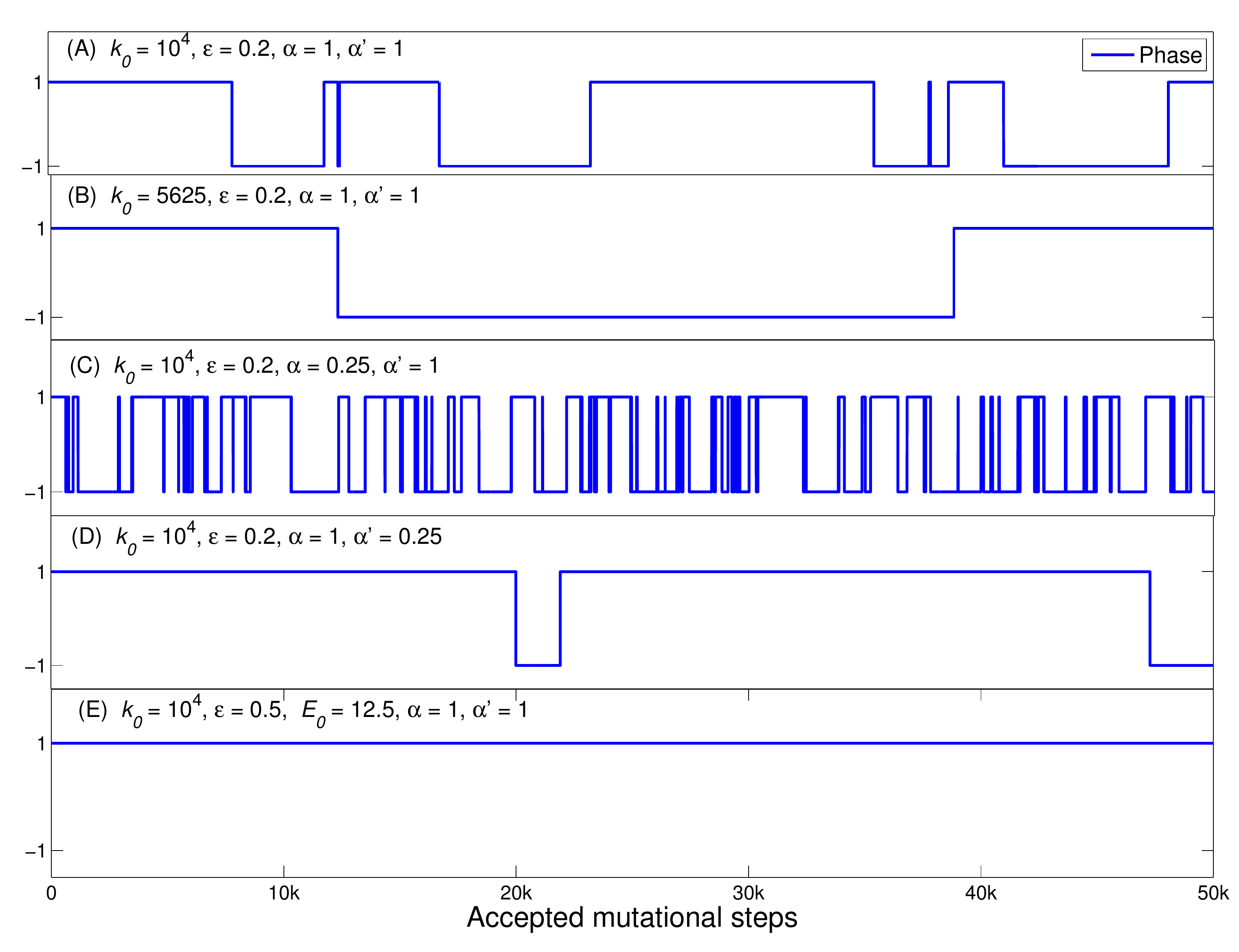}
\caption{Depiction of the temporal evolution of phases of 2 activator 1 deactivator systems, where a value of +1 indicates Phase 1 and -1 indicates Phase 2. The plots are obtained by varying the model parameters, the interaction energy between hydrophobic residues $\epsilon$, the rate constant $k_0$, and background activation and deactivation rates $\alpha$ and $\alpha^{'}$.}
\label{Robust2}
\end{figure}

To confirm the robustness of our results with respect to the choice of model parameters, we used different values of the interaction energy between hydrophobic residues $\epsilon$, the rate constant $k_0$, and background activation and deactivation rates for the 2 activator, 1 deactivator systems. Qualitatively we find that all our results hold as long as the maximum interaction energy $E_{\rm max} = N \epsilon$, does not substantially exceed the threshold energy $E_0$, that is, provided $E_{\rm max} - E_{0} < 1$, where energy is expressed in units of $k_{\rm B}T$; otherwise the reaction rates become saturated at their maximum values over a large fraction of sequence space. Keeping all other parameters fixed, if we vary $k_0$, we find that the system remains in a particular phase longer for smaller $k_0$. The duration of phases is shorter for $k_0 = 10^4$ (Fig. \ref{Robust2}~A) compared to $k_0=5625$ (Fig. \ref{Robust2}~B). For a smaller value of background activation rate ($\alpha = 0.25$) for both the activators and deactivator, we find that the duration of phases becomes shorter, i.e., the system switches between Phase 1 and Phase 2 more rapidly (Fig. \ref{Robust2} C). On the other hand for a smaller value of background deactivation rate ($\beta = 0.25$) the phase duration becomes longer (Fig.~\ref{Robust2}~D).  The quantities relative PVR, phase-shift, and PC1 continue to be correlated with the phase of the system (Phase 1 or Phase 2) (Fig.~\ref{Robust1}). Alternately, if we increase the value of $\epsilon$ while maintaining $E_{0} = N \epsilon$, we find a significant increase in the phase duration, so that for $\epsilon = 0.5$ we hardly see  transitions from one phase to the other (Fig.~\ref{Robust2}~E). It is worth noting that if $E_{\rm max} = N \epsilon$ is allowed to be much greater than $E_{0}$, then physically the concentrations of the intermediates corresponding to enzymes bound to targets should become significant, and one would need to explicitly account for their concentrations in the rate equations, thus increasing their complexity. A systematic investigation of this case lies outside the scope of the current paper.

\clearpage

\section{Figure details}
As different size data sets were used for each figure, we provide the details here. The duration of each single simulation is $10^5$ accepted mutational steps.
 
	{\bf Fig. 1A}: A single run for $10^5$ accepted mutational steps.

	{\bf Fig. 1B}: 4000 data points used to produce the histogram for each starting sequence.

	{\bf Fig. 2A-D}: Single run for $10^5$ accepted mutational steps.

	{\bf Fig. 2E}: 100 simulations each running for $10^5$ accepted mutational steps.

	{\bf Fig. 3A}: Approximately 6000 data points generated by running 150 simulations each running for $10^5$ accepted mutational steps.

	{\bf Fig. 3B}: Approximately $10^5$ data points generated by running 150 simulations each running for $10^5$ accepted mutational steps.

	{\bf Fig. 4A}: A single run for $10^5$ accepted mutational steps.

	{\bf Fig. 4B}: Approximately $10^6$ data points generated by running 80 simulations each running for $10^5$ accepted mutational steps.

	{\bf Fig. 4C-D}: Single run for $2\times10^5$ accepted mutational steps.


\begin{thebibliography}{30}
\bibitem{footnote1}
Since we find that states where both activators are individually inessential are very rare, approximately $0.001\%$ of the total number of oscillatory states, we ignore such states for the purposes of the figure.
\label{footnote1}
\end{thebibliography}
\end{document}